\documentclass[trackchanges]{aastex701}

\usepackage{units}

\usepackage{amsmath}
   \usepackage[normalem]{ulem}
\definecolor{mypink}{rgb}{0.858, 0.188, 0.478}
\definecolor{mygrey}{rgb}{0.55, 0.68, 0.55}

 \newcommand{\s}{\nobreak\hspace{.11em}\nobreak}

 \newcommand{\be}{\begin{equation}}
 \newcommand{\ee}{\end{equation}}
 \newcommand{\ba}{\begin{eqnarray}}
 \newcommand{\ea}{\end{eqnarray}}
 \newcommand{\bs}{\begin{subequations}}
 \newcommand{\es}{\end{subequations}}

\begin{document}

    \title{Synchronisation of a tidal binary by inward orbital migration. The case of Pluto and Charon}

\author[orcid=0000-0003-1249-9622,gname=Michael,sname=Efroimsky]{Michael Efroimsky}
\affiliation{US Naval Observatory, Washington DC 20392 USA}
\email[show]{michael.efroimsky@gmail.com}
\correspondingauthor{Michael Efroimsky}

\author[orcid=0000-0002-6779-3848,gname=Michaela,sname=Walterov\'{a}]{Michaela Walterov\'{a}}
\affiliation{Department of Geophysics, Faculty of Mathematics and Physics, Charles University, Prague 12116 Czech Republic}
\email[show]{kanovami@gmail.com}

\author[0000-0002-8723-3434,gname='Yeva',sname=Gevorgyan]{Yeva Gevorgyan}
\affiliation{King Abdullah University of Science and Technology, Thuwal 23955-6900 Saudi Arabia}
\email[show]{yeva@ime.usp.br}

\author[0000-0002-2306-2576,gname=Amirhossein,sname=Bagheri]{Amirhossein Bagheri}
\affiliation{California Institute of Technology, Pasadena CA 91125 USA}
\email{abagheri@caltech.edu}

\author[orcid=0000-0003-2336-7887,gname=Valeri,sname=Makarov]{Valeri V. Makarov}
\affiliation{US Naval Observatory, Washington DC 20392 USA}
\email[show]{valeri.makarov@gmail.com}

\author[orcid=0000-0002-7787-4836,gname=Amir,sname=Khan]{Amir Khan}
\affiliation{Institute of Geochemistry and Petrology, ETH Z\"{u}rich, CH-8092 Z\"urich Switzerland}
\email[show]{amir.khan@eaps.ethz.ch}

 \begin{abstract}
It is usually taken for granted that mutual synchronisation of a tidal two-body system is attained through tidal recession, assuming the reduced Hill sphere is not reached. However, synchronisation can be achieved also via tidal approach, provided the Roche limit is not crossed. For each of the two scenarios, we derive the condition under which the evolving synchronicity radius catches up with the tidally evolving orbit. We consider the two scenarios for the Pluto-Charon system and examine the impact-origin hypothesis of Charon's formation against capture. {Based on geophysical evidence, we propose that capture appears more likely. Motivated by this conclusion, we investigate both analytically and numerically the capture scenario, wherein the orbital evolution of Charon starts at a larger distance than present and undergoes tidal descent. We also consider the possibility that Pluto’s initial prograde spin underwent a reversal by a tidally approaching retrograde Charon. Depending on the initial conditions, we observe temporary locking of Charon into higher spin-orbit resonances (3:2 to 7:2) during the first $\unit[0.5]{Myr}$ of the system's evolution. Owing to a greater initial separation between the partners, the power dissipated in each of them turns out to be much lower than in the case of tidal recession of bodies of the same internal structure.
The greater initial separation also results in lower tidal stress, which  may explain the absence of tidally generated fracture patterns.}
\end{abstract}


\section{Introduction}
\label{Introduction}

The most thoroughly studied trans-Neptunian objects (TNOs), Pluto and Charon, are often thought to have formed from the remnants of two differentiated or partially differentiated impacting bodies
\citep{Canup2005, Stern, Desch2015, DeschNeveu2017}. Within such impact models, the orbit of Charon is predicted to be initially very close to Pluto, and to subsequently expand until reaching its current synchronous state, as a result of tidal dissipation within both partners, with Pluto's rotation simultaneously slowing down \citep{2020A&A...644A..94C, Renaud_2021, Amirs}. Yet, some predictions from giant-impact simulations are at odds with recent observations. The differences include the mass ratio unusual for a planet-satellite system, the apparent similarity in bulk structure, the absence of tidally driven fractures on Charon, the lack of a fossil bulge on Pluto, and the retrograde rotation of Pluto \citep{Stern2018}.\\

Had Charon been formed by a giant impact, it could be expected to be less massive yet richer in ice. Charon, however, carries as much as one eighth of Pluto's mass, and appears to have a bulk structure not very different from Pluto. Specifically, it only shows a slightly higher ice-to-rock ratio. For a composition of water ice plus anhydrous rock, following solar abundances, Pluto and Charon contain, correspondingly, about 2/3 and 3/5 rock by mass \citep{McKinnon2017}.
Both Charon's large relative mass and appreciable rock fraction are difficult to explain within a regular graze-and-merge scenario without special means. These means were developed by \citet{Canup2005} who postulated the collision to be (1) oblique, (2) low-velocity, and (3) by an undifferentiated impactor with mass exceeding Pluto's and Charon's total mass by 0.3. Each of these three requirements is potentially possible. However, their combination, while not impossible, is less probable than each of them separately.

   Along similar lines, \citet{Arakawa_2019} found that to form Charon-like large satellites with a moderately high ice-to-rock mass fraction, one needs an  impact angle larger than $60^{\circ}$ (though this depends on the pre-impact configuration) and a low impact velocity (typically, exceeding the escape velocity by a factor of $1$ to $1.2\s$). By distinction from \citet{Canup2005},  the model developed by \citet{Arakawa_2019} permits formation of the Pluto–Charon system through a collision of two differentiated progenitors, as illustrated in Figure 1 in {\it Ibid}. However, for differentiated progenitors, the range of initial conditions producing a satellite-planet mass ratio as high as that of Charon to Pluto is narrow.
The rarity of these initial conditions may be consistent with the exceptional status of Charon and Pluto in the sense of their mass ratio, among the satellites of Kuiper-Belt dwarf planets \citep{Canup2005, Arakawa_2019}.

An alternative formation mechanism that would explain the similar densities of Pluto and Charon was recently proposed by \citet{Denton2025}. In their ``kiss-and-capture'' scenario, with the effect of material strength included, Charon is collisionally captured by Pluto; that is, the two bodies become temporarily connected, and upon detachment form a binary system with compositions and interior structures only slightly affected by the merger~---~a development different from that emerging in the giant-impact simulations by, e.g.,~\citet{Canup2005}.

Most scenarios for the post-impact orbital evolution of Charon suggest an internal ocean and a high initial eccentricity. As pointed out in \citet{Rhoden} and references therein, these conditions would have produced tidal stresses comparable to other tidally fractured satellites, such as Europa and Enceladus. No correlation, however, has been found in {\it Ibid.} between the observed fracture orientations and those predicted to emerge under eccentricity-caused tidal stress. To explain the mismatch, the authors had to hypothesise (a) that Charon's orbit circularised before its ocean froze, and (b) that during the circularisation process, either tidal stresses alone were too weak to fracture the surface or the primordial fractures were removed by subsequent resurfacing.

Any impact scenario implies that Pluto's initial rotation was much faster than now. Fast initial rotation, however, would have made Pluto's figure manifestly oblate. After synchronisation, the oblateness would relax, albeit not completely, and a residual fossil bulge would be expected to remain. No such rotation bulge was discovered by {\it New Horizons} in limb profiles, and the data analysis indicated that the shapes of Pluto and Charon are spherical to within 0.6\% and 0.5\%, respectively \citep{Nimmo,Nimmo2021}. The consequences of Pluto's fossil rotational oblateness being below the precision level of {\it New Horizons}'s imagery and occultation measurements were addressed quantitatively by \citet{plutocharon}. This study concluded that a low oblateness can be put in agreement with the impact hypothesis only by adopting some special restrictions on Pluto's structure\,---\,e.g. on its lithosphere thickness, under the assumption that a thin lithosphere could make Pluto soft and prevent it from sustaining a permanent, Iapetus-like oblateness. An alternative explanation for the lack of a fossil equatorial bulge would be the existence of a hypothetical internal ocean \citep{Nimmo2021}. This option was considered in detail by \citet{Amirs}, who combined a comprehensive tidal model with a parameterised thermal convection model developed for icy worlds, and found that a present-day ocean on Pluto emerges under a variety of initial conditions, whereas an initial ocean on Charon would have most likely solidified because of its smaller size.

The observations summarised above suggest that Pluto and Charon:
\begin{itemize}
 \item[(1)] may not have been created by a destructive giant
  impact~---~while an oblique low-velocity impact \citep{Canup2005, Arakawa_2019} or a collisional capture \citep{Denton2025} remain options capable of retaining the original ice-to-rock fraction;
\item[(2)] were able to relax to their present-day shape upon despinning, either due to a weak ice shell or due to the presence of a subsurface ocean over some part of their history \citep{plutocharon};
\item[(3)] have experienced tidal stresses weaker than the stresses induced by ocean solidification~\citep{Rhoden}.
\end{itemize}

In the aforementioned studies of the tidal evolution of the Pluto-Charon system, mutual synchronisation was assumed to have been attained via Charon's recession \citep{Robuchon, cheng_etal14,
Amirs,Denton2025}. However, as was pointed out, e.g.,~in \citet{pathways}, synchronisation of a planet-moon pair via tidal descent is, generally, another valid option. As was later proposed in \citet{synchronisation}, an initially prograde Pluto could have captured Charon at a large angle with respect to the ecliptic~---~whereafter Charon synchronised Pluto by tidally approaching it and changing its rotation direction. While synchronisation of Pluto and Charon during inward tidal migration from a larger initial separation does not automatically exclude the possibility of Charon's formation by an impact, it makes capture a more likely option---particularly in the light of Pluto's retrograde spin. Moreover, the lack of tidal fractures may also be simpler to explain by the capture scenario. This would imply that the existing fractures were formed during the freezing process that either preceded capture or succeeded the system's full tidal evolution. After capture, tidally driven fractures may or may not have appeared, depending on whether the capture took place below or above the present orbit. Tidal descent from a larger initial separation would cause much weaker stresses than tidal ascent from a closer distance, and would be much slower. Thus, the absence of tidal fractures supports synchronisation in falling.\\

The possibility of capture was demonstrated by \citet{Agnor}, who modelled three-body gravitational encounters of a solitary planet with a binary asteroid
 comprising two near-equal-mass partners. Upon dissociation of the binary, one of the partners gets captured by the planet.  While several TNOs (2007 OR10, Haumea, Quaoar, Makemake) are within a factor of a few Charon masses and have moons \citep[Table S1]{Arakawa_2019}, these moons are considerably smaller than their hosts.  Also, while close-mass TNO pairs are numerous, their masses are smaller than Charon's (e.g., both Ceto and Phorcys are $\sim$70 times lighter than Charon).  Nonetheless, based on comparisons to other large objects and formation models, \citet{Proudfoot} demonstrated that formation of large near-equal-mass pairs is possible. Produced by gravitational collapse of pebble clouds \citep{Robinson},
  these pairs survive implantation into the dynamically excited Kuiper-belt environment \citep{Nesvorny}. Recently, \cite{Williams_and_Zugger} demonstrated that the binary-exchange capture scenario also works for planets less massive than Neptune and for asteroid binaries with masses differing by a factor of four or more.\\


A very different mechanism, chaos-assisted capture, was described in \citet{Venus} and references therein. That mechanism favours capture of retrograde moons over prograde. It was specifically demonstrated in \citet{Venus} that a so-captured retrograde moon may have reversed Venus' initially prograde spin as it migrated inward.\\

To further the study of captured binaries, we investigate post-capture dynamics, including a possible reversal of Pluto's initially prograde spin by a tidally approaching retrograde Charon, which follows from the angular momentum conservation. We also find that Charon could have been temporarily captured into higher spin-orbit resonances during the early stages of its orbital evolution. As mentioned above, owing to an initially larger separation of Charon from Pluto, tidal stresses in both bodies are much lower. This provides an explanation for the lack of tidally generated fracture patterns on Charon, which in the case of the opposite scenario -- an impact followed by outward migration -- is not easily accounted for.

\section{Simplified analytical treatment of dynamics}\label{dynamics}


\subsection{Assumptions}

We begin with a number of assumptions. First of all, we limit our consideration to a near-coplanar configuration:
 \begin{itemize}
 \item[(a)] The moon's orbit is near-equatorial, with its inclination $i$ relative to the planet's equator being small enough to permit the omission of $O(i^2)$.
 \item[(b)]
 The moon’s obliquity $i_{\rm{m}}$ relative to its orbit is small enough to allow for the neglect of $O(i_{\rm{m}}^2)$ terms.
 \end{itemize}

Although the present rotation of Pluto and orbital motion of Charon are retrograde with respect to the prograde direction observed in most cases in the solar system, we use, for mathematical convenience, an opposite convention.  We regard Pluto's present rotation and Charon's present orbit as ``prograde'', and set the mean motion to be positive: $n>0$. If Charon's initial orbital motion was concordant with or opposite to Pluto's initial spin, Pluto's initial rotation rate, $\Omega$, is regarded positive ($\Omega>0$) or negative ($\Omega<0$), correspondingly.

\vspace{2mm}
To simplify the analytical model further, we invoke two additional assumptions:

\begin{itemize}
\item[(c)] The spin angular momentum of the moon can be neglected.
\end{itemize}
\noindent
As demonstrated in Appendix \ref{A1}, Charon's spin brings to dynamical equations an input of order $10^{-3}$ for eccentricities up to $0.85$, after Charon's synchronisation. The input from a nonsynchronous Charon is, of course, larger. So this simplification is good after Charon is slowed down, but is less precise at the start, when Charon is swiftly rotating.

\begin{itemize}
\item[(d)] The eccentricity is sufficiently low for the $O(e^2)$ terms to be dropped.
\end{itemize}
\noindent
This assumption simplifies the analytical treatment presented below, but imposes a limitation on its precision.  The reason for this is that the leading-order, in the powers of $e$, expressions for the orbital elements' rates and the partners' spin evolution may cause noticeable errors already for $e\gtrsim 0.2$ \citep{Amirs,Renaud_2021}. With limited obliquities and eccentricity assumed, our analytical development provides only a qualitative description.
\vspace{1mm}

While all of the above items (a -- d) are relied upon in the simplified analytic model developed in Sections \ref{synchronisa}, \ref{catch} and \ref{histories}, only the first two items, (a) and (b), will be retained in the numerical model constructed in Section \ref{numerics}. Items (c) and (d) will be foregone in the numerical runs, so the evolution of both the spin angular momentum of the moon and the eccentricity could be modelled consistently.

 \begin{table*}

 \caption{Summary of properties of the Pluto–Charon system. Variables without an assigned value are calculated or varied in the numerical runs.}
 \label{description}
 ~\\
 \begin{tabular}{@{}llll@{}}
 \hline\\
   \vspace{1mm}
 Variable & ~~~~~Value & Description & Reference \\
 \hline \\

 \vspace{1.4mm}

 $  M                 $   &                $ 1.302\times 10^{22}$ kg                   &  Pluto's mass          &   \citet{brozovic2024}                      \\[2pt]

 \vspace{1.4mm}

 $  R                $   &               $1.1883\times 10^6$ m                    &  Pluto's mean radius   &   \cite{Nimmo}                  \\[2pt]

 \vspace{1.4mm}

 $ C $   &     & Pluto's maximal moment of inertia     &                  \\[2pt]

 \vspace{1.4mm}

 $  \xi\,\equiv\,\frac{\textstyle C}{\textstyle M\s R^{\s 2}}
  $      &                $0.306$                   &  Pluto's moment of inertia factor (MoIF)              &   Section 4.3      \\[2pt]

 $  M_{\rm{m}}  $   &  $ 1.5897  \times 10^{ 21} $ kg  &  Charon's mass        &  \citet{brozovic2024}  \\[2pt]

 \vspace{1.4mm}

 $  R_{\rm{m}}  $   &  $ 0.6060 \times 10^{  6} $  m    &  Charon's  mean radius & \cite{Nimmo}   \\[2pt]

 \vspace{1.4mm}

$ C_{\rm{m}} $   &     & Charon's maximal moment of inertia     &                  \\[2pt]

 \vspace{1.4mm}

$  \xi_{\rm{m}}\,\equiv\,\frac{\textstyle C_{\rm{m}}}{\textstyle M_{\rm{m}}\s R_{\rm{m}}^{\s 2}}~~~
  $      &                $0.315$                   &  Charon's moment of inertia factor (MoIF)              &   Section 4.3      \\[2pt]

 \vspace{1.4mm}

 $ k_2
 $  &                            & Pluto's Love number          &     \\[2pt]

 \vspace{1.4mm}

 $  Q                       $   &                            & Pluto's  quality factor &          \\[2pt]

 \vspace{1.4mm}

 $  K_2$  &
 & Pluto's quality function &  Equation (\ref{formula})                   \\[2pt]

\vspace{1.4mm}

 $ {k}_{{\rm{m},}\, 2}
 $  &                            & Charon's Love number          &                   \\[2pt]

 \vspace{1.4mm}

 $  Q_{ \rm{m}}                       $   &                            & Charon's  quality factor &          \\[2pt]

 \vspace{1.4mm}

 $  {K}_{{\rm{m},}\, 2} $  &
 & Charon's quality function &  Equation (\ref{formula})                  \\[2pt]

\vspace{1.4mm}

 $  \Omega $  &    & Pluto's rotation rate &                    \\[2pt]

 \vspace{1.4mm}

  $ \Omega_{\rm{p}}  $  & $1.1567 \times 10^{-5}~\mbox{rad/s}$   & Pluto's rotation rate, present value &    \citet{brozovic2024}               \\[2pt]

 \vspace{1.4mm}

  $  \Omega_{\rm{m}} $  &    & Charon's rotation rate &                    \\[2pt]

 \vspace{1.4mm}

 $ a $   &     & Charon's semimajor axis     &                  \\[2pt]

 \vspace{1.4mm}

 $ a_{\rm{p}} $   &  $19.596 \times 10^6~\mbox{m}$   & Charon's semimajor axis, present value     &  \citet{brozovic2024}                \\[2pt]

 \vspace{1.4mm}

 $ n $   &     & Charon's mean motion     &                  \\[2pt]

\vspace{1.4mm}

 $  e
  $      &                                   &  Charon's eccentricity              &         \\[2pt]

 \vspace{1.4mm}

 $  e_{\rm{p}}
  $      &   $0.0002$                                &  Charon's eccentricity, present value              & \citet{brozovic2024}        \\[2pt]

 \vspace{1.4mm}

  $  i
  $      &   $0$                                &  Charon's inclination              & \citet{brozovic2024}        \\[2pt]

 \hline
 \end{tabular}
 ~\\ \vspace{4mm}
  \end{table*}

 \subsection{Synchronisation scenarios}
\label{synchronisa}

Two tidal synchronisation scenarios are possible:
\begin{itemize}
\item[] Scenario 1. ~Mutual synchronism is attained through orbital recession accompanied by the despinning\\
$\phantom{\qquad\;\quad\qquad}$  of both bodies.
\item[] Scenario 2. ~Mutual synchronism is attained through orbital decay. This can occur for a moon in a\\
$\phantom{\qquad\;\quad\qquad}$ prograde orbit below the synchronous radius, or for a moon in a retrograde orbit.
\end{itemize}

For Scenario 1 to occur, the moon must be formed or captured above the planet's synchronous radius at that time,\,\footnote{~A less likely but not impossible option is formation or capture of a rapidly spinning moon slightly below synchronism. The tides in the moon can overpower those in the planet and push the moon above the synchronous radius. For this, the moon must initially possess spin angular momentum sufficient to overcome the angular momentum difference between the initial and synchronous orbits.} and its eccentricity on approach to synchronism should not be  too high.\,\footnote{~A moon, which has already synchronised its spin but still retains a high eccentricity, can cross the synchronous orbit from above, if the planet-caused tides in the moon overpower the moon-generated tides in the planet. Likely, the case of Phobos \citep{Bagheri}.}

In Scenario 2, the tides on the planet act to shrink the orbit in both cases~---~whether the moon is in a prograde orbit below synchronous, or in a retrograde orbit.\,\footnote{~A retrograde moon is tidally descending, except in a special case of a slowly rotating planet and an extremely quickly rotating moon.\label{exception}}

After the spin of the moon is synchronised with the mean motion, the tides in the moon are working (for $e\neq 0$) towards orbital decay, within both scenarios.

As the moon approaches the planet, the planet's rotation is either accelerated, for a prograde moon, or decelerated, for a retrograde moon. In the latter situation, the planet may change the direction of its rotation in the inertial frame. If the planet's angular acceleration rate $\dot{\Omega}$ at some point exceeds the rate $\dot{n}$ of the moon's orbital decay, the planet's rotation can be equalised with the moon's orbital motion before the moon reaches the Roche limit.

 \subsection{
 The condition of catching up
 \label{catch}}

 In both above scenarios, the changing synchronous radius has to catch up with the evolving orbit. For small eccentricity and obliquities, and neglecting the spin angular momentum of the moon, this process can be described by the equation
 \ba
 \frac{\stackrel{\bf\centerdot}{\Omega}}{\stackrel{\bf\centerdot}{n}}\,=\,\frac{1}{3\,\xi}\,\frac{M_{\rm{m}}}{M+ M_{\rm{m}}}\,\left(\frac{a}{R}\right)^2+\,O(e^2)
 \label{3ksi.eq}
 \ea
 derived in Appendix \ref{A1} from the angular-momentum conservation law.
 Here $\Omega$ is the planet's spin rate,
 $R$ is its radius, $\xi$ is its moment of inertia factor (the polar moment of inertia divided by $MR^2$), while $M$ and $M_{\rm{m}}$ are the planet's and moon's masses, respectively. As customary, $a$ and $n$ are the semimajor axis and mean motion, respectively.
 (These and other notations are summarised in Table 1.)
 \vspace{4 mm}

  \subsubsection*{Catching up in Scenario 1}

 A prograde moon's initial orbit is above synchronous, implying $\s\Omega - n > 0\s$. The end-state is $\s\Omega -n = 0\s$, and is attained if the inequality
$\s\frac{\textstyle d}{\textstyle dt}\s(\Omega - {n}) < 0\s$ is fulfilled \textcolor{black}{from some moment in time onward~---~though not necessarily from the start.}
 In this scenario,
 both angular accelerations ${\stackrel{\bf\centerdot}{\Omega\s}}$ and ${{\stackrel{\bf\centerdot}{n\s}}}$ are negative ($n$ itself being positive, as stipulated~above).

\textcolor{black}{
 As will be explained in Section \ref{histories}, see Figure \ref{Figure_1} and equation (\ref{flip}), the moon's recession is faster than the expansion of the synchronous radius
 (equivalently, $\s{\textstyle d}\s(\Omega - {n})/{\textstyle dt} > 0\s$)
 for as long as the planet-moon separation is below a critical value denoted with $a_{\rm\s flip}$. As soon as the moon crosses that value of $a$, its ascent becomes slower than the growth of the synchronous radius (that is to say, $\s{\textstyle d}\s(\Omega - {n})/{\textstyle dt} < 0\s$, see equation \ref{larger}).
 }

 To summarise,
 \bs
 \ba
 \nonumber\\
 \left.
 \begin{array}{lll}
 \left.
 \begin{array}{lll}
 \mbox{Initial state:}\qquad \Omega - n >0  \vspace{4mm}\\
 \mbox{End state:} \qquad\;\,\; \Omega - n =0
 \end{array}
 \right\}
 \phantom{\;}      \Longrightarrow    \phantom{\;} \frac{\textstyle d}{\textstyle dt}\s(\Omega - {n}) < 0
 \textcolor{black}{~,~~\mbox{after $a$
 surpasses $a_{_{\rm\s{flip}}}\s$
  }}
 \\
 ~\\
 \phantom{\;\;}\mbox{Prograde, above-synchronous:} \quad  {\stackrel{\bf\centerdot}{\Omega\s}}\,<\,0\;,\;\;\;{{\stackrel{\bf\centerdot}{n\s}}}\,<\,0
 \end{array}
 \right\}
         \nonumber\\  \nonumber\\  \nonumber\\
 \phantom{\;\;}    \Longrightarrow   \phantom{\;\;\;} {\stackrel{\bf\centerdot}{\Omega\s}}\,<\,{\stackrel{\bf\centerdot}{n\s}}\,<\,0
 \textcolor{black}{~,~~\mbox{after $a$
 surpasses $a_{_{\rm\s{flip}}}\s$
  }}
  \;\;,
 \qquad
         \qquad\qquad\qquad
 \label{aa}
 \ea
 which may also be written as
 \ba
 |\s{{\stackrel{\bf\centerdot}{\Omega\s}}}\s |\,>\,|\s{{\stackrel{\bf\centerdot}{n\s}}}\s |
 \textcolor{black}{~,~~\mbox{after $a$
 surpasses $a_{_{\rm\s{flip}}}\s$
  }}
  \;\;.
 \label{bb}
 \ea
 \label{inequa1}
 \es
 Thus the absolute value of the despinning rate of the planet should be higher than the absolute rate of the moon's orbital slowdown.

 Combined with equation (\ref{3ksi.eq}), inequality (\ref{bb}) yields
 \ba
 3\,\xi\,\frac{M\s+\s M_{\rm{m}}}{M_{\rm{m}}}\,\left(\frac{R}{a}\right)^2\s <\,1
 \,\;,
 \label{}
 \ea
 which simplifies to
 \ba
 a\,>\,a_{{\rm\s  flip}}\,=\,R\,\sqrt{3\,\xi\,\frac{M\s+\s M_{\rm{m}}}{M_{\rm{m}}}\,}\textcolor{black}\,\;.
 \label{larger}
 \ea
 Since in this scenario the moon is receding, the semimajor axis is growing in time. Hence, once condition (\ref{larger}) is fulfilled for some value of $a$, its fulfilment is warranted until either synchronism or the reduced Hill radius is reached.

As emphasised in \citet{pathways}, it is not  necessary for this inequality to be satisfied from the start. If the initial configuration is such that the moon is above synchronism but still close to the planet, inequality (\ref{larger}) may, in principle, be violated.  In this case, violated will be requirement (\ref{inequa1}), and $\s|\s{{\stackrel{\bf\centerdot}{n\s}}}\s |\s$ will be larger than $\s|\s{{\stackrel{\bf\centerdot}{\Omega\s}}}\s |\s$.  So the moon's orbital ascent will be progressing faster than the planet's spin-down.
The expansion rate $\s{{\stackrel{\bf\centerdot}{a\s}}}\s$ of the orbit will be exceeding the rate
$\s{{\stackrel{\bf\centerdot}{a\s}}_{\rm{s}}}\s$ of the synchronous radius $a_{\rm{s}}\s$, and the synchronous radius will be falling behind the receding moon.\,\footnote{~From the expression $a_{\rm{s}} = \left(\s G\s(M+M_{\rm{m}})/\Omega^{\s 2}\right)^{1/3}$,
  \,it follows that
 $$
 \frac{\textstyle
 \,{{\stackrel{\bf\centerdot}{a\s}}_{\rm{s}}}\,
 }{\textstyle
 {{\stackrel{\bf\centerdot}{a\s}}\;}}\,=\,
 \frac{\textstyle
 {{\,\stackrel{\bf\centerdot}{\Omega\s}\,}}
 }{\textstyle
 {{\stackrel{\bf\centerdot}{n\s}}}
 }\,
 \left(
 \frac{\textstyle \,n\,}{\textstyle \Omega}
 \right)^{5/3}\;\;.
 $$
 }
 If this sequence is never reversed, the moon will leave the reduced Hill sphere and will be lost.\,\footnote{~The inequality ${\textstyle
 {{\stackrel{\bf\centerdot}{a\s}}}} > {\textstyle
 {{\stackrel{\bf\centerdot}{a\s}}_{\rm{s}}}
 }$ is never reversed if $a_{\rm flip}= R\sqrt{3\xi (M+M_{\rm{m}})/M_{\rm{m}}}$ exceeds the reduced Hill radius. Certainly, not the case~for Pluto and Charon. Pluto's reduced Hill radius for prograde and retrograte moons, correspondingly, is $3.75\times 10^9\,$m  $\approx 192\, a_p\,$ and $7.12\times 10^9\,$m  $\approx 364\, a_p\s$, \,where $a_p$ is the present value of Charon's semimajor axis. See, e.g. equations (2) and (3) in \citet{pathways}.\label{foot}} However, as $a$ is growing, the situation may change. In this case, inequality (\ref{larger}) begins to be satisfied at a certain instant~--~and remains satisfied thereafter. This gives the moon a chance to synchronise the planet's rotation before leaving the reduced Hill sphere.

  \subsubsection*{Catching up in Scenario 2}

  A prograde moon's initial orbit is below synchronous, {\s}with $\,\Omega\s -\s n\s <\s 0\,$.  The end-state with $\s\Omega -n = 0\s$ is then attained by keeping
 $\s\frac{\textstyle d}{\textstyle dt}\s(\Omega - {n}) >0\s$, both angular accelerations ${\stackrel{\bf\centerdot}{\Omega\s}}$ and ${{\stackrel{\bf\centerdot}{n\s}}}$ staying positive.
 Concisely,
 \ba
 \nonumber\\
 \nonumber
 \left.
 \begin{array}{lll}
 \left.
 \begin{array}{lll}
 \mbox{Initial state:}\qquad \Omega - n <0  \vspace{4mm}\\
 \mbox{End state:} \qquad\;\,\; \Omega - n =0
 \end{array}
 \right\}
 \phantom{\;}      \Longrightarrow    \phantom{\;} \frac{\textstyle d}{\textstyle dt}\s(\Omega - {n}) >0\\
 \nonumber\\
 \phantom{\;\;}\mbox{Prograde, below-synchronous:} \quad  {\stackrel{\bf\centerdot}{\Omega\s}}\,>\,0\;,\;\;\;{{\stackrel{\bf\centerdot}{n\s}}}\,>\,0
 \end{array}
 \right\}
 ~\\
 \nonumber\\
 \nonumber\\
 \phantom{\;\;}    \Longrightarrow   \phantom{\;\;\;} {\stackrel{\bf\centerdot}{\Omega\s}}\,>\,{{\stackrel{\bf\centerdot}{n\s}}}\,>\,0
 \;\;,
 \qquad\qquad\qquad\qquad
 \label{inequa2}
 ~\\
 \nonumber
 \ea
 Mathematically, the processing of inequality (\ref{inequa2}) is identical to the afore-performed processing of inequality (\ref{inequa1}). Combining inequality (\ref{inequa2}) with the angular-momentum conservation law (\ref{3ksi.eq}), we again arrive at
 \ba
 a\,>\,a_{{\rm\s  flip}}\,=\,R\,\sqrt{3\,\xi\,\frac{M\s+\s M_{\rm{m}}}{M_{\rm{m}}}\,}\,\;,
 \label{large}
 \ea
 though on this occasion the physical meaning of this inequality is different: it sets the initial separation sufficiently large to ensure that the descending moon has enough time to synchronise the planet.
 Moreover, as pointed out in \citet{pathways}, within Scenario 2 this inequality must be satisfied {\it from the start}, lest the dynamics follow Scenario 1. Indeed, if this inequality~---~and therefore also inequality (\ref{inequa2})~---~are initially violated, the moon will begin receding, not descending.

 {The system comes to an equilibrium}
 at the moment of synchronisation, provided this synchronisation takes place above the Roche radius (otherwise the moon disintegrates, and synchronism is never achieved). So, while in Scenario 1 it was necessary to keep the value of $\s R\s\sqrt{(M+M_{\rm{m}})/M_{\rm{m}}}\s$ below the reduced Hill radius (see Footnote \ref{foot}), in Scenario~2 this value must exceed the Roche radius.

The formalism describing Scenario 2 remains valid for an initially retrograde moon, one captured on an orbit directed oppositely to the initial spin direction of the planet. For this moon, a synchronous orbit does not exist~---~or may, when convenient, be assumed  infinite. Tides in the planet are working to pull such a moon down. If sufficiently massive, a tidally descending moon may reverse the planet's rotation. Thereafter, the moon may perish below the Roche limit, leaving the planet in a retrograde rotation mode~---~which may be Venus' case \citep{Venus}. Alternatively, the moon may synchronise the planet above the Roche limit. In this situation, both the final orbit of the moon and the resulting rotation of the planet are retrograde~---~which may be the case of Pluto and Charon \citep{synchronisation}.

 For mathematical convenience, we set the mean motion of Charon positive, $n>0$. Pluto's initial rotation rate $\Omega$ can be of either sign.  Specifically, it was negative ($\Omega<0$) in the case when Charon's initial orbital motion was opposite to Pluto's initial spin. This warrants that, initially, $\,\Omega\s -\s n\s <\s 0\,$. Therefore, the case with opposite signs of $n$ and the initial $\Omega$ is described by the Scenario 2 formalism. We emphasise that a negative initial $\Omega$ implies that Pluto was initially ``retrograde'' only with respect to Charon's initial mean motion $n$. With respect to the solar system's standard prograde direction, things were opposite: Pluto's initial spin was prograde, while Charon's initial mean motion was retrograde.

\section{Dynamical Histories}
\label{histories}

  \subsection{Charon's synchronous orbit:  Analysis of tidal stability}
 \label{instabi}

  As shown in Appendix \ref{A1}, for small eccentricity and obliquities, and in neglect of the spin angular momentum of the moon,
  the angular momentum conservation law assumes the form of equation (\ref{3ksi.eq}).
  At low eccentricities, integration of this equation \citep{synchronisation} results in a linear relationship between $\sqrt{a}$ and $\Omega$\s:
 \ba
 \Omega\s=\s-\s X\s \sqrt{ a}\,+\,C\,\;,
 \label{5}
 \label{Omega}
 \ea
 where the slope is given by
 \ba
 X\s=\s\frac{M_{\rm{m}}}{\xi}\,\sqrt{\frac{G}{M\s +\s M_{\rm{m}}}\,}\s R^{\s -2}
 \;\,,
 \label{X}
 \label{4}
 \ea
 while the constant is linked to the initial conditions through
 \ba
 C\,=\,X\s\sqrt{a(t_0)}\s+\s\Omega(t_0)\;\,.
 \label{C}
 \ea
  \textcolor{black}{The steepness of the synchronisation track as defined by equation (\ref{X}) is strongly dependent on the size of the planet, $X\propto R^{-7/2}$. Hence, rotation of small planets undergoes much more drastic changes in the course of tidal evolution.}

 A system, whose evolution began at a point $(\sqrt{a(t_0)\s}\s,\,\Omega(t_0)\s)\s$, cannot arrive at a synchronous spin-orbit state unless its trajectory intersects with the set of all synchronous states available for the given masses. Recalling that the synchronicity condition reads as $\Omega = n$, we find that in the axes $(\sqrt{a}\s,\,\Omega)$ the set of synchronous states is depicted with the cubical hyperbola
 \ba
 \Omega=\sqrt{G\,(M\s +\s M_{\rm{m}})\,}\,(\!\sqrt{a})^{-3}\;.
 \label{6}
 \label{OmegaS}
 \ea
\begin{center}
\begin{figure}	\includegraphics[width=.8\columnwidth]{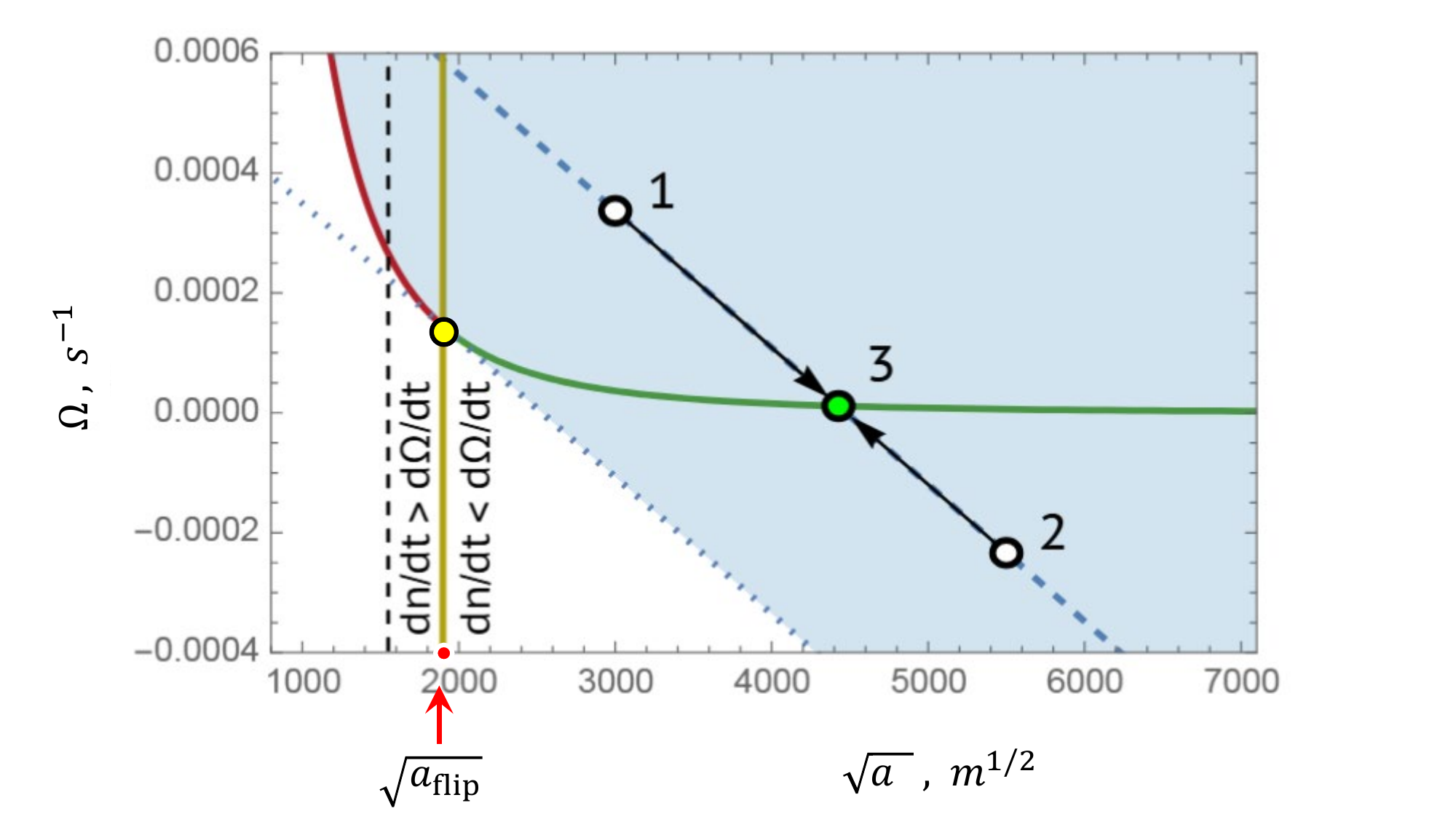}
   \caption{
   The cubical hyperbola represents
   all synchronous states given by equation (\ref{6}), with the green-coloured branch indicating long-term stable equilibria, and the red-coloured part showing the loci of intrinsically unstable equilibria.
   The blue dashed and dotted lines represent two evolution tracks described by equation (\ref{5}). The track on the right results in a stable capture of Pluto (the green disk). The one on the left traverses the critical point (the yellow disk). The vertical dashed line corresponds to the Roche limit. The light blue-shaded area of the plot marks all initial conditions that may result in long-term synchronisation of Pluto. The states marked with white disks and labelled ``1'' and ``2'' are possible initial conditions that could lead to the current state of the system, which is shown with the green disk and labelled ``3''.
   The $1\rightarrow 3$ dynamical history corresponds to Scenario 1, with a tidally receding moon  \textcolor{blue}{(equations \ref{inequa1} - \ref{larger})}. The $2\rightarrow 3$ track illustrates Scenario 2, with a moon tidally approaching the planet  \textcolor{blue}{(equations \ref{inequa2} - \ref{large})}.
   {The semimajor axis value corresponding to the critical point (the yellow disk) is named $a_{\rm\s flip}$ because in Scenario~1 the tidal recession of a moon is faster than the growth of the synchronous radius, insofar as $a<a_{\rm\s flip}$. On crossing $a_{\rm\s flip}$, the growth of the synchronous radius becomes faster, and it catches up with the moon at point ``3'' (the green disk).
   Crossing of $a_{\rm\s flip}$ can happen also when synchronisation is not achieved, i.e., when the track misses the hyperbola below the tangent point, going from right to left and up. The difference \textcolor{blue}{
   ${\stackrel{\bf\centerdot}{\Omega\s}}\,-\,{{\stackrel{\bf\centerdot}{n\s}}}$
   } between the derivatives
   still changes sign there, outside the scenarios of synchronisation.
   }
}
    \label{Figure_1}
    \end{figure}
\end{center}
  As can be seen in Figure~\ref{Figure_1}, for a fixed pair of mass values, $M$ and $M_{\rm{m}}$, the number of available synchronous spin-orbit states can be zero, or one, or two~---~depending on the system's initial state $(\sqrt{a(t_0)\s}\s,\,\Omega(t_0)\s)$. The stability of these synchronous states was explored by different methods in \citet{Darwin1879}, \citet{Counselman}, \citet{Hut_1980}, and \citet{synchronisation}. It was determined that when a physical trajectory (a straight line in Figure~\ref{Figure_1}) has two common points with the cubical curve, the left point is unstable, while the right point is long-term stable.
Mathematically, the difference between the left and right crossing points is that $d\Omega/dn <1$ holds at the former, $d\Omega/dn >1$ at the latter.
  \textcolor{black}{
 The dashed track in Figure ~\ref{Figure_1} shares with the cubical curve only one point, marked with a yellow disk and corresponding to a semimajor axis value $a_{\rm\s flip}$. Since at this point ${\stackrel{\bf\centerdot}{\Omega}}={\stackrel{\bf\centerdot}{n}} $, we obtain from equation (\ref{3ksi.eq}) that
 \ba
 a_{\rm\s flip}\s=\,R\,\sqrt{3\,\xi\,\frac{M\s+\s M_{\rm{m}}}{M_{\rm{m}}}\,}\textcolor{black}\,\;.
 \label{flip}
 \ea
 }
Inserting Pluto's and Charon's parameters into equation (\ref{3ksi.eq}), and setting $\s 3\s\xi=1\s$, we obtain:
  \ba
  \frac{\textstyle d\Omega}{\textstyle dn}\s=\s
  \frac{\stackrel{\bf\centerdot}{\Omega}}{\stackrel{\bf\centerdot}{n}}\,=\,29.51\;\gg\;1\;\,,
  \label{}
  \ea
  which ensures tidal stability. The Pluto-Charon system is residing at the right intersection point in Figure~\ref{Figure_1}.


  \subsection{Analytical approximation of the initial conditions}
 \label{analytics}

We aim to find a set of initial conditions $\s(a_0\s,\,\Omega_0)\s$ that result in dynamical trajectories terminating at the end state $\s(a\s,\,\Omega)\s$ satisfying $\Omega\s=\s n\s=$
 $\sqrt{ {\textstyle G\s(M+M_{\rm{m}})} / {\textstyle a^{3}}\,}\,$. We identify this end state with the present-day state, and endow both $a$ and $\Omega=n$ with the present values for the Pluto-Charon system, $a_{\rm{p}}$ and $\Omega_{\rm{p}}$. Using an analytical approximation, we can obtain approximate initial conditions $\s(a_0\s,\,\Omega_0)\s$ for trajectories terminating in $\s(a_{\rm{p}}\s,\,\Omega_{\rm{p}})\s$.

Taking into account equations (\ref{X}  - \ref{C}), a dynamical history (\ref{Omega}) can be expressed as
 \ba
 \sqrt{a_0}\s=\s\sqrt{a}\,\left[1\,+\,B^{-1}\s\left(1\,-\,\frac{\Omega_0}{\Omega}\right)\s   \right]\,\;,
 \label{dddd}
 \ea
 where
 \ba
 B\s\equiv\s\xi^{-1}\,\frac{M_{\rm{m}}}{M\s+\s M_{\rm{m}}}\;\left( \frac{a}{R}  \right)^{\s 2}\;=\;88.52\,\;,\quad\mbox{for}\quad\xi = {1}/{3}\;\,.
 \label{BBB}
 \ea
  We fix the values of $a$ and $\Omega$ at the instant of synchronisation (these may be identified with the present-day values $a_{\rm{p}}$ and $\Omega_{\rm{p}}\s$).
 For a value $\Omega_0$ of the planet's rotation rate at the moment of formation or capture of the moon, equation (\ref{dddd}) will then give us the separation $a_0$ at that same moment.

 Equation (\ref{dddd}) describes both the synchronisation through tidal recession of the moon ($a_0 < a$ $\s\Longrightarrow\s$ $\Omega_0 > \Omega $) and the synchronisation through its tidal descent ($a_0 > a$ $\s\Longrightarrow\s$ $\Omega_0 < \Omega $). In the latter case, $\Omega_0$ may be of either sign.

Evidently, the initial orbit must have resided between the Roche limit and the reduced Hill radius (see Footnote \ref{foot} for its value).
 Also, each dynamical history converging to a stable state in Figure \ref{Figure_1} naturally satisfies the condition that $\Omega_0$ exceeds $n_0=\sqrt{G(M+M_{\rm{m}})/a_0^3\,}$, for a tidally falling moon (the $2\rightarrow 3$ evolution track in Figure \ref{Figure_1}), or that $\Omega_0$ is smaller than $n_0$, for a tidally ascending moon (the $1\rightarrow 3$ history in Figure \ref{Figure_1}).
 No such condition applies to a retrograde moon, because in realistic situations it is always tidally descending, see Footnote \ref{exception}.

Based on expression (\ref{3ksi.eq}), our analytical formulae are approximate for two reasons. First, we neglected the spin angular momentum of Charon~---~which simplification brings in an error of only $\sim 0.1\%$ (see Appendix \ref{A1}) for an already synchronised Charon's spin, but may generate a larger error prior to Charon's synchronisation. For $e\gtrsim 0.2\s$, the precision of expression (\ref{3ksi.eq}) is further limited by the neglect of $O(e^2)$ terms.\,\footnote{~The inaccuracy grows rapidly for eccentricities higher than $0.2$ because of a very slow convergence of series over the powers of $e$ in the tidal theory. As was demonstrated for realistic models in, e.g. \citet{Amirs} and \citet{Renaud_2021}, order-of-magnitude differences begin to occur for $e > 0.6\s$.
 \label{Foot}} The analytical approximation given by equation (\ref{dddd}) is intended for small eccentricities and a slow initial spin of Charon, and can be used to constrain the initial orbit parameters for numerical integration.  Table \ref{tab:eq13} demonstrates the accuracy of the approximation across different initial eccentricity values ($0 \leq e_0 \leq 0.4$) and Pluto's initial spin rates (${\Omega_0}_{\s m}/\Omega_{\rm{p}} \in \{0.8, 2, -5\}$), showing where approximation (\ref{dddd}) remains reliable, and where we should use the angular momentum conservation law in an exact form (see equation \ref{eq:quartic_b0} derived below). For an initial eccentricity of $e_0 = 0.1$ we have an error of $\s\,\sim 1\%$. For $e_0=0.2$, the precision is $\s\,\sim 4\%$. For $e_0$ approaching $0.4$, the error becomes too large for approximation (\ref{dddd}) to be used.

\begin{table}[h]
\centering
\begin{tabular}{|c|c|c|c|c|}
\hline
$e_0$ & ${\Omega_{\rm{m}0}}/\Omega_{\rm{p}}=\Omega_0/\Omega_{\rm{p}}$ & $a_{0}/a_{\rm{p}}$ (Eq. \ref{route}) & $a_{0}/a_{\rm{p}}$ (Eq. \ref{dddd}) & $\left(a_{0}^{\rm{(A1)}}-a_{0}^{(\rm{12})}\right)/a_{0}^{(\rm{12})}$\\
\hline
$0$ & $0.8$  & $1.00467$ & $1.00452$ & $0.014\,\%$ \\
\hline
$0.1$ & $0.8$  & $1.01481$  & $1.00452$ & $1.024\,\%$\\
\hline
$0.2$ & $0.8$  & $1.04653$  & $1.00452$ & $4.181\,\%$\\
\hline
$0.3$ & $0.8$  & $1.10403$ & $1.00452$ & $9.906\,\%$\\
\hline
$0.4$ & $0.8$  & $1.19603$ & $1.00452$ & $19.064\,\%$\\
\hline
$0$ & $2$  & $0.976834$ & $0.977534$ & $-0.072\,\%$ \\
\hline
$0.1$ & $2$  & $0.986701$  & $0.977534$ & $0.938\,\%$\\
\hline
$0.2$ & $2$  & $1.01753$  & $0.977534$ & $4.092\,\%$\\
\hline
$0.3$ & $2$  & $1.07344$  & $0.977534$ & $9.811\,\%$\\
\hline
$0.4$ & $2$  & $1.1629$ & $0.977534$ & $18.962\,\%$\\
\hline
$e_0$ & $-{\Omega_{\rm{m}0}}/\Omega_{\rm{p}}=\Omega_0/\Omega_{\rm{p}}$ & $a_{0}/a_{\rm{p}}$ (Eq. \ref{route}) & $a_{0}/a_{\rm{p}}$ (Eq. \ref{dddd}) & $\left(a_{0}^{\rm{(A1)}}-a_{0}^{(\rm{12})}\right)/a_{0}^{(\rm{12})}$\\
\hline
$0$ & $-5$  & $1.13706$ & $1.14016$ & $-0.272\,\%$ \\
\hline
$0.1$ & $-5$  & $1.14854$ & $1.14016$  & $0.736\,\%$ \\
\hline
$0.2$ & $-5$  & $1.18443$ & $1.14016$ & $3.884\,\%$ \\
\hline
$0.3$ & $-5$  & $1.24951$ & $1.14016$ & $9.591\,\%$ \\
\hline
$0.4$ & $-5$  & $1.35364$ & $1.14016$ & $18.724\,\%$\\
\hline
\end{tabular}
\caption{Comparison of the semimajor axis' value obtained from the approximate equation (\ref{dddd}) with the exact value resulting from the conservation of total angular momentum, (\ref{route}), for both prograde and retrograde Pluto.}
\label{tab:eq13}
\end{table}

 \section{Numerical simulations. General plan and parameters}
 \label{numerics}

\subsection{Orbital evolution model}

To explore the orbital evolution of the Pluto-Charon system numerically, we adopt a tidal evolution model based on a corrected version of the Darwin-Kaula formalism \citep{darwin80,kaula61,Kaula,BoueEfroimsky2019}. Within this formalism, the tidal potential as well as the evolution equations for all Keplerian elements and for the spin states are written in the form of a sum over multiple tidal modes $\{l,m,p,q\}$, arising due to the mutual tidal loading of the two bodies in question. The tidal modes perceived by Pluto due to the gravitational action of Charon are:

\be
    \omega_{lmpq} = (l-2p+q)\s n - m\, \Omega \; .
\ee\\
Correspondingly, the tidal modes perceived by Charon perturbed by Pluto are:

\be
     {\omega_{\rm{m,\,}}}_{lmpq} = (l-2p+q)\s n - m\, \Omega_{\rm{m}}\; ,
\ee\\
where $l$, $m$, $p$, $q$ are integers. The absolute values of tidal modes, $|\omega_{lmpq}|$ and $|  {\omega_{\rm{m\,}}}_{lmpq}|$, are the actual physical frequencies of tidal waves
\citep[Section 4.3]{EfroimskyMakarov13}.

For a non-synchronous planet hosting a non-synchronous satellite, in neglect of physical libration and after averaging over the orbital period and the period of apsidal precession, the evolution equations for the semimajor axis $a$, the eccentricity $e$, the spin rates $\Omega$, $\Omega_{\rm{m}}$, and the dissipated energies $E$, $E_{\rm{m}}$ read as \citep{BoueEfroimsky2019,efroimsky2014}:

\be
  \frac{{\rm{d}}a}{{\rm{d}}t} = -2\, a\, n \sum_{lmpq} \frac{(l-m)!}{(l+m)!}\, (2-\delta_{m0})\; \mathcal{G}_{lpq}^2(e)\; (l-2p+q)\; \mathcal{S}_{lmpq}\; , \label{eq:dadt}
\ee

\be
  \frac{{\rm{d}}e}{{\rm{d}}t} = -\frac{n}{e} \sum_{lmpq} \frac{(l-m)!}{(l+m)!}\, (2-\delta_{m0})\; \mathcal{G}_{lpq}^2(e)\; \left[(1-e^2) (l-2p+q) - \sqrt{1-e^2} (l-2p)\right]\; \mathcal{S}_{lmpq}\; , \label{eq:dedt}
\ee

\be
    \frac{{\rm{d}}\Omega}{{\rm{d}}t} = \frac{G M_{\rm{m}}^2}{a\, \xi\, M R^2} \sum_{lmpq} m \frac{(l-m)!}{(l+m)!}\, (2-\delta_{m0})\; \left(\frac{R}{a}\right)^{2l+1} \mathcal{F}_{lmp}^{2}(i)\; \mathcal{G}_{lpq}^2(e)\; K_l(\omega_{lmpq})\; , \label{eq:dOdt}
\ee

\be
    \frac{{\rm{d}}\Omega_{\rm{m}}}{{\rm{d}}t} = \frac{G M^2}{a\, \xi_{\rm{m}}\, M_{\rm{m}} R_{\rm{m}}^2} \sum_{lmpq} m \frac{(l-m)!}{(l+m)!}\, (2-\delta_{m0})\; \left(\frac{R_{\rm{m}}}{a}\right)^{2l+1} \mathcal{F}_{lmp}^{2}(i_{\rm{m}})\; \mathcal{G}_{lpq}^2(e)\;
    {K_{\rm{m,\s}}}_l(  {\omega_{\rm{m,\,}}}_{lmpq})\; , \label{eq:dOmdt}
\ee

\be
    \frac{{\rm{d}}E}{{\rm{d}}t} = \frac{G M_{\rm{m}}^2}{a} \sum_{lmpq} \frac{(l-m)!}{(l+m)!}\, (2-\delta_{m0})\; \left(\frac{R}{a}\right)^{2l+1} \mathcal{F}_{lmp}^{2}(i)\; \mathcal{G}_{lpq}^2(e)\; |\omega_{lmpq}|\, K_l(\omega_{lmpq})\; , \label{eq:dEEdt}
\ee

\be
    \frac{{\rm{d}}E_{\rm{m}}}{{\rm{d}}t} = \frac{G M_{\rm{m}}^2}{a} \sum_{lmpq} \frac{(l-m)!}{(l+m)!}\, (2-\delta_{m0})\; \left(\frac{R_{\rm{m}}}{a}\right)^{2l+1} \mathcal{F}_{lmp}^{2}(i_{\rm{m}})\; \mathcal{G}_{lpq}^2(e)\; |\omega_{{\rm{m,\,}}lmpq}|\, {K_{\rm{m,\,}}}_l(  {\omega_{\rm{m,\,}}}_{lmpq})\; , \label{eq:dEEmdt}
\ee
where we used a shortened notation for the sum over all tidal modes,

\ba
    \sum_{lmpq}\equiv\sum_{l=2}^{+\infty} \sum_{m=0}^{l} \sum_{p=0}^{l} \sum_{q=-\infty}^{+\infty}\; ,
\label{}
\ea
~\\
and a function $\mathcal{S}_{lmpq}$ containing the tidal response of both companions,

\begin{equation}
    \mathcal{S}_{lmpq} = \left(\frac{R}{a}\right)^{2l+1} \frac{M_{\rm{m}}}{M} \mathcal{F}_{lmp}^{2}(i) K_l(\omega_{lmpq}) + \left(\frac{R_{\rm{m}}}{a}\right)^{2l+1} \frac{M}{M_{\rm{m}}} \mathcal{F}_{lmp}^{2}(i_{\rm{m}}) {K_{\rm{m,\,}}}_l(  {\omega_{\rm{m,\s}}}_{lmpq}
    )\; .
\end{equation}\\
In the above equations, $K_l$ and ${K_{\rm{m\s}}}_l$ denote the planet's and the moon's tidal quality functions, such that
\ba
|K_l| = \frac{k_l}{Q_l}\qquad \mbox{and} \qquad |{K_{\rm{m,\,}}}_l|=\s\frac{{k_{\rm{m,\,}}}_l}{{Q_{\rm{m,\,}}}_l}\;\;.
\label{formula}
\ea
Additionally, $\mathcal{G}_{lpq}$ are the eccentricity functions and $\mathcal{F}_{lmp}$ are the inclination functions \citep{kaula61,allan65}. The ``inclination'' in the context of the latter functions is the angle between a celestial body's equator and the orbital plane of the perturber. For Pluto perturbed by Charon, this angle is equal to the inclination $i$. For Charon perturbed by Pluto, this angle is equal to Charon's obliquity $i_{\rm{m}}$. Since we are focusing on a two-dimensional problem, we set both $i$ and $i_{\rm{m}}$ zero,
and model the evolution of the semimajor axis and eccentricity, ignoring the other orbital elements. To calculate the eccentricity functions, which are a subset of Hansen coefficients, we employ the Von Zeipel-Andoyer method, as described by \citet{izsak64} and \citet{cherniack72}.

The Darwin-Kaula series is expanded for $l=2$ and $q\in[-6,6]$, i.e., we retain harmonic modes up to the 12th power in eccentricity. The retaining of higher-order terms and the corresponding modes is necessary, because Charon's initial orbit around Pluto may have been highly eccentric \citep[e.g.,][]{Canup2005}. A similarly detailed approach, with expansion up to $e^{20}$, was employed in the past to understand, among other things, the origin of the Martian moons \citep{Bagheri}, icy moons, and TNOs \citep{Amirs, Renaud_2021, arakawa_eta21}.

The evolution equations (\ref{eq:dadt} - \ref{eq:dEEmdt}) are implemented as in \citet{walterova20}, with two amendments. First, the updated model includes the tidal response of both partners. Second, ahead of the numerical runs, the values of the tidal quality functions are precalculated for the prescribed interior structures at multiple frequencies (between $10^{-10}$ and $\unit[10^{-3}]{\,rad\;s^{-1}}$).
The runs are carried out using a fourth-order predictor-corrector integration scheme with a variable step size \citep[e.g.,][]{ralston65}.
At each step of the integration, the tidal quality functions for all relevant tidal modes are estimated by interpolation from the precalculated and tabulated values.

\subsection{Calculation of the tidal quality functions}

Assuming that the planets are described as spheres consisting of layers with homogeneous properties, the tidal deformation and tidally-induced changes in the external gravitational potential can be estimated from normal mode theory \citep[e.g.,][]{sabadini04,tobie05b}. Within this approach, the continuity equation, the equation of motion, and the Poisson equation for the gravity field are expanded into spherical harmonics and rewritten as a set of ordinary differential equations for six radially dependent functions $\{y_1,...,y_6\}$. The set of equations has six linearly independent solutions. The coefficients of their combination can then be determined from the boundary conditions at the surface and all interior interfaces, and from the requirement of regularity in the centre.

We seek the coefficients of the linear combination by solving a system of linear equations for the boundary conditions. We also reconstruct the displacements and additional gravity potential at the surface of the planet and calculate the tidal Love numbers from their definition \citep{sabadini04}.

By default, all layers in our implementation are treated as viscoelastic, and all interfaces are solid-solid. The introduction of liquid layers is done by prescribing a negligibly small shear modulus and a low viscosity. This approach, preferred for its generality when treating thermally evolving bodies \citep{walterova20}, was also tested against an alternative implementation  of the equations presented in \citet{sabadini04} with prescribed solid-liquid interfaces (and assuming only static tides in the liquid layers), which yielded identical results. Since the layers are considered viscoelastic, the planet's reaction to tidal loading is frequency-dependent and the periodic deformation also results in tidal dissipation. The resulting Love numbers of viscoelastic bodies are then generally complex, and the tidal quality functions $K_l(\omega_{lmpq})$ and ${K_{\rm{m,\,}}}_l(\omega_{{\rm{m,\,}}lmpq})$ can be obtained as the negative imaginary part of the complex potential Love numbers at the relevant frequencies. Both the absolute values of the complex tidal Love numbers and the quality functions of our Pluto and Charon models are depicted in Figure \ref{fig:k2Q}, illustrating the frequency dependence of these tidal parameters at successive stages of the orbital evolution of the system.

 \subsection{Initial conditions and model parameters}
 \label{initial}

If the final pair of values $(a_{\rm{p}}\s,\,\Omega_{\rm{p}})$ is known (where the subscript `p' means {\it present}), a corresponding history $(\s a(t)\s,\,\Omega(t)\s)$ will terminate at this final point. The analytical relation (\ref{dddd}) provides an approximate link between the end point $(a\s,\,\Omega)\s=\s(a_{\rm{p}}\s,\,\Omega_{\rm{p}})$ and a starting point $(a_0\s,\,\Omega_0)$, with an error originating from two sources: one from our neglect of the spin angular momentum of the moon (see Appendix~\ref{A1}),  another from our omission of the $O(e^2)$ term in equation (\ref{3ksi.eq}).

An exact value of the initial semimajor axis, $a_0$, can be obtained from the conservation of total angular momentum, expressed by equation (\ref{to}). Using a substitution $s=\sqrt{a}$, and assuming that the initial eccentricity $e_0$ and spin-orbit ratios $\Omega_0/n_0$, ${\Omega_{\rm{m}}}_0/n_0$ are known, and recalling that the present-day state corresponds to a full synchronisation with a near-zero eccentricity, equation (\ref{to}) can be rewritten in the form of a quartic equation for the initial value of $s$, denoted as $s_0$:

\ba
s_0^4 \left(\frac{M\; M_{\rm{m}}}{M+M_{\rm{m}}} \sqrt{1-e_0^2}\right) - s_0^3 \left[\frac{M\; M_{\rm{m}}}{M+M_{\rm{m}}} s_{\rm{p}} + \left(C + C_{\rm{m}}\right) s_{\rm{p}}^{-3}\right] + \frac{1}{n_0} \left(C\s \Omega_0 + C_{\rm{m}}\s
{ \Omega_{\rm{m}}}_0\right) = 0\;\,. \label{eq:quartic_b0}
\ea\\
While this equation, in principle,  can be solved analytically, we calculate the solution iteratively, starting from the present-day value of $s\equiv s_{\rm{p}}$.

To model a dynamical scenario within which a rapidly spinning Charon-sized body is gravitationally captured by a rapidly spinning Pluto, we use canonically accepted asteroid rotation periods.  Pluto’s initial rotation period is taken to lie in the range $[5.4,\; 14]$ hours. This interval is consistent with the rotation periods of Kuiper Belt objects larger than 100 km, which range from about 5.3 h (Vesta) to 14 h (2001 QG298), see \citet{Canup2005}. Charon's initial rotation period is either chosen to be the same as for Pluto or two times smaller or greater. The initial eccentricity values are taken from the range $[0.1, 0.5]$. For each set of parameters, we calculate the initial semimajor axis from equation (\ref{eq:quartic_b0}), which ensures that we arrive at the present state of the Pluto-Charon system. An overview of all study cases and the initial parameters used can be found in Table \ref{tab:initial_parameters}.\\

We note that due to the limited range of tested initial eccentricities, all initial semimajor axis values satisfying the conservation of total angular momentum fall within the interval from $1.5$ to $2\; a_{\rm{p}}$. As a consequence, the total tidally generated heat in the examples presented in Section \ref{runs} is limited by the small total change in semi-major axis. According to \citet{Agnor} and \citet{Venus}, post-capture orbits tend to be highly eccentric; an initial eccentricity of, e.g., $e_0=0.8$, would then require $a_0/a_{\rm{p}}\approx4-5$. A larger initial semimajor axis results in a greater amount of the total tidally-generated heat, which is mainly dissipated towards the end of the orbital evolution. Nevertheless, due to a rapid tidal circularisation of highly eccentric orbits, the evolution timescale in large-$a_0$ cases does not increase considerably when compared to smaller $a_0$.\\

The solid parts of both bodies are characterised by the viscoelastic Andrade model \citep{Andrade62,Efroimsky2012,gev2020}. This model includes the elastic response as well as viscous and transient creep and is therefore favoured over the Maxwell model \citep[e.g.][]{castillo2011tidal, bagheri_etal19, pou2022tidal}. For the Andrade parameters $\alpha$ and $\zeta$, which specify the frequency dependence and the strength of transient creep, respectively, we set $\zeta=1$ and $\alpha=0.2$. \textcolor{black}{Since we focus on the orbital and rotational evolution of the Pluto-Charon system, we endow both partners with a simplified interior structure comprising three homogeneous layers. The properties of these layers, such as the viscosity, rigidity, or density, then correspond to the average (or effective tidal) values. Specifically, for the viscosity of ice, we adopt a value of $\unit[10^{14}]{Pa\;s}$, and we also test the effect of increasing the ice shell's effective  tidal viscosity to $\unit[10^{16}]{Pa\;s}$. It should be mentioned that neither the structure of Pluto's ice shell nor its viscosity value are fully understood. The choice of a low average viscosity ($\unit[10^{14}]{Pa\;s}$) approximately corresponds to a warm and convecting ice shell in contact with a subsurface ocean. Although it has been previously argued that an early onset of convection in the ice shell leads to efficient heat loss and full ocean freezing \citep{Robuchon,nimmo_etal16}, a low ice viscosity might still be compatible with the present-day existence of a subsurface ocean if the ice shell contains an insulating basal layer of clathrates \citep{Kamata2019}.} The \textcolor{black}{other} physical properties of water ice are adopted from \citet{Amirs, bagheri_etal25a} and \citet{petricca_etal24}. For Charon, we scale the \textcolor{black}{layer} thicknesses with the relative size of Charon compared to Pluto.

\begin{figure}
    \centering
    \includegraphics[width=0.9\textwidth]{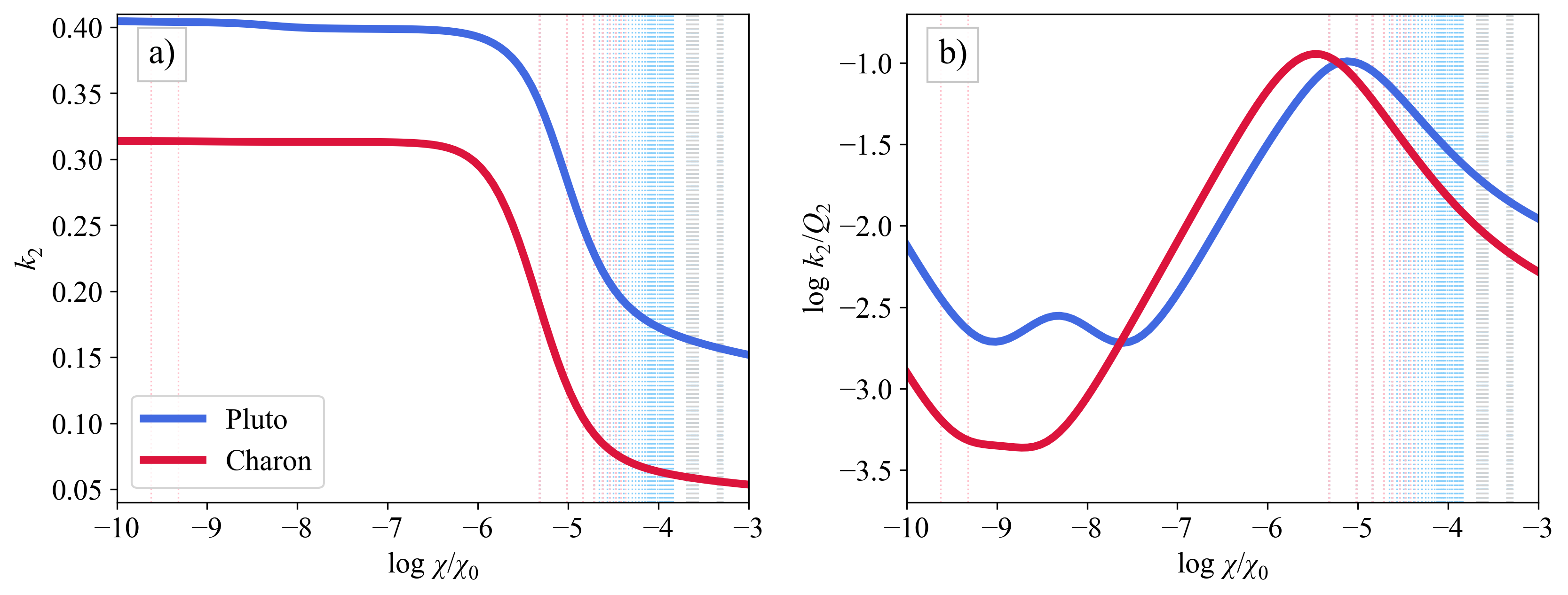}
    \caption{\textcolor{black}{(a) The absolute value of the tidal potential Love number $k_2$ and (b) the logarithm of the tidal quality function $|K_2|=k_2/Q_2$ as a function of the physical frequency $\chi$ for the three-layered models of Pluto and Charon presented in Table \ref{tab:interior_props}. The reference frequency is $\chi_0=\unit[1]{rad\,s^{-1}}$. The vertical lines indicate the frequencies relevant at the beginning of the evolution (gray dotted line, similar values for both bodies) and for the spin-orbital configuration after $\unit[1]{Myr}$ (light blue and pink dotted lines for Pluto and Charon, respectively) in the case with $\Omega_0/n_0=50$ (Case 1, Table \ref{tab:initial_parameters}). They serve as an illustration of the typical tidal frequencies rather than an overview of all frequencies used.}}
    \label{fig:k2Q}
\end{figure}

All orbital evolution runs in our study are performed under the assumption that there are no structural changes to the planetary interiors with time. Evidence suggests that present-day Pluto retains a subsurface ocean, while Charon’s has solidified \citep{nimmo_etal16, Amirs}. However, as our computations pertain to a distant past, we assume that both bodies harboured subsurface oceans at that time. Since an ocean’s presence increases the tidal Love number, the partners' $k_2$ values are much closer to one another than those predicted for the contemporary Pluto and Charon \textcolor{black}{(see Figure \ref{fig:k2Q})}. The interior profiles considered here are summarised in Table \ref{tab:interior_props}.

The densities of the silicate parts of Pluto and Charon are specifically adjusted to yield the correct masses, as listed in Table \ref{description}. For the three-layered models from Table \ref{tab:interior_props}, we also obtain the moment of inertia factors $\xi=0.306$ and $\xi_{\rm{m}}=0.315$, needed to calculate the evolution of spin rates. In all considered histories, we observe a change in Pluto's rotation direction, making it retrograde as currently observed.

\begin{table}
\begin{center}
\begin{tabular}{l l l l l l}
\hline

 $P_{0}$ [h] \hspace{0.7cm} & $|\Omega_{0}/\Omega_{\rm{p}}|$ \hspace{0.7cm} & $|\Omega_{0}/n_{0}|$ \hspace{0.7cm} & $\Omega_{\rm{m}0}/\Omega_{0}$ \hspace{0.7cm} & $a_{0}/a_{\rm{p}}$ \hspace{0.7cm} & $e_0$ \\
\hline
\multicolumn{6}{c}{\textbf{Case 1:} $e_0=0.4$, $\Omega_{\rm{m}0}/\Omega_{0}=-1$} \\
\multicolumn{6}{c}{Effect of initial spin rates, depicted in Figure \ref{fig:1to1}.}\\
\hline
-5.40 & 28.40 & 80 & & 1.995 & \\
-6.14 & 25.00 & 65 & & 1.891 & \\
-7.26 & 21.12 & 50 & -1 & 1.777 & 0.4 \\
-9.26 & 16.56 & 35 & & 1.647 & \\
-14.00 & 10.96 & 20 & & 1.495 & \\
\hline
\multicolumn{6}{c}{\textbf{Case 2:} $\Omega_0/n_0=50$, $\Omega_{\rm{m}0}/\Omega_{0}=-1$} \\
\multicolumn{6}{c}{Effect of initial eccentricity, depicted in Figure \ref{fig:e_variation}.}\\
\hline
-6.20 & 24.75 & & & 1.598 & 0.1 \\
-6.38 & 24.04 & & & 1.630 & 0.2 \\
-6.72 & 22.83 & 50 & -1 & 1.686 & 0.3 \\
-7.26 & 21.12 & & & 1.777 & 0.4 \\
-8.14 & 18.85 & & & 1.917 & 0.5 \\
\hline
\multicolumn{6}{c}{\textbf{Case 3:} $e_0=0.4$, $\Omega_0/n_0=35$} \\
\multicolumn{6}{c}{Effect of initial spin rate ratio, depicted in Figure \ref{fig:1to1_negative}.}\\
\hline
-9.53 & 16.09 & & 2 & 1.679 & \\
-9.44 & 16.24 & 35 & 1 & 1.668 & 0.4 \\
-9.40 & 16.32 & & 0.5 & 1.663 & \\
\hline
\end{tabular} \\[0.3em]
\caption{Initial parameters leading to the current synchronous state of the Pluto-Charon system. Here, $n_0$ is the initial mean motion, $P_0$ is Pluto's initial rotation period, and $\Omega_0$ and $\Omega_{\rm{m}0}$ are the initial spin rates of Pluto and Charon, respectively. The present rotation rate of both partners (equal to their present mean motion) is denoted with $\Omega_{\rm{p}}$.
}
\label{tab:initial_parameters}
\end{center}
\end{table}

\section{Results of numerical simulations}
\label{runs}

\begin{table}
\begin{center}
\begin{tabular}{lcc }
\hline
\textbf{Parameter/unit}                                   & \textbf{Value}   \\ \hline
\multicolumn{1}{l}{Ice layer}      &       &           \\
\multicolumn{1}{l}{\quad Shear modulus ~(GPa)   }                 & 3.0      \\
\multicolumn{1}{l}{ \quad  Density~(g/$\rm cm^3$)}       & 0.92  \\
\multicolumn{1}{l}{ \quad  Viscosity ~(Pa.s)}       & 10$^{14}$  \\
\multicolumn{1}{l}{ \quad  Thickness (Pluto)~(km)}    & 208  \\
\multicolumn{1}{l}{ \quad  Thickness (Charon)~(km)}    & 106  \\
\multicolumn{1}{l}{Water layer} &                   &        \\
\multicolumn{1}{l}{\quad Shear modulus ~(GPa)   }               & 0       \\
\multicolumn{1}{l}{\quad Density~(g/$\rm cm^3$) }             & 1.0       \\
\multicolumn{1}{l}{ \quad  Viscosity ~(Pa.s)}       & 10$^{-3}$  \\
\multicolumn{1}{l}{ \quad  Thickness (Pluto)~(km)}       & 120  \\
\multicolumn{1}{l}{ \quad  Thickness (Charon)~(km)}       & 60  \\
\multicolumn{1}{l}{Silicate layer} &        &        \\
\multicolumn{1}{l}{\quad Shear modulus  ~(GPa)  }                   & 60      \\
\multicolumn{1}{l}{\quad Density (Pluto)~(g/$\rm cm^3$)}                   &  3.3   \\
\multicolumn{1}{l}{\quad Density (Charon)~(g/$\rm cm^3$)}                   &  2.9   \\
\multicolumn{1}{l}{\quad Viscosity ~(Pa.s)}                   &  10$^{20}$  \\
\multicolumn{1}{l}{\quad Radius (Pluto) ~(km)}      &  860   \\
\multicolumn{1}{l}{\quad Radius (Charon) ~(km)}      &  440   \\
\hline \hline
\end{tabular}
\caption{Parameter values for the interior models of Pluto and Charon. Values for ice and water are adopted from \citet{Amirs}, while those for silicate are taken from \citet{McKinnon2017}. The uppermost $\unit[6]{km}$ for Charon and $\unit[8]{km}$ for Pluto are considered elastic.}
\label{tab:interior_props}
\end{center}
\end{table}

\begin{figure}[h]
  \includegraphics[width=\textwidth]{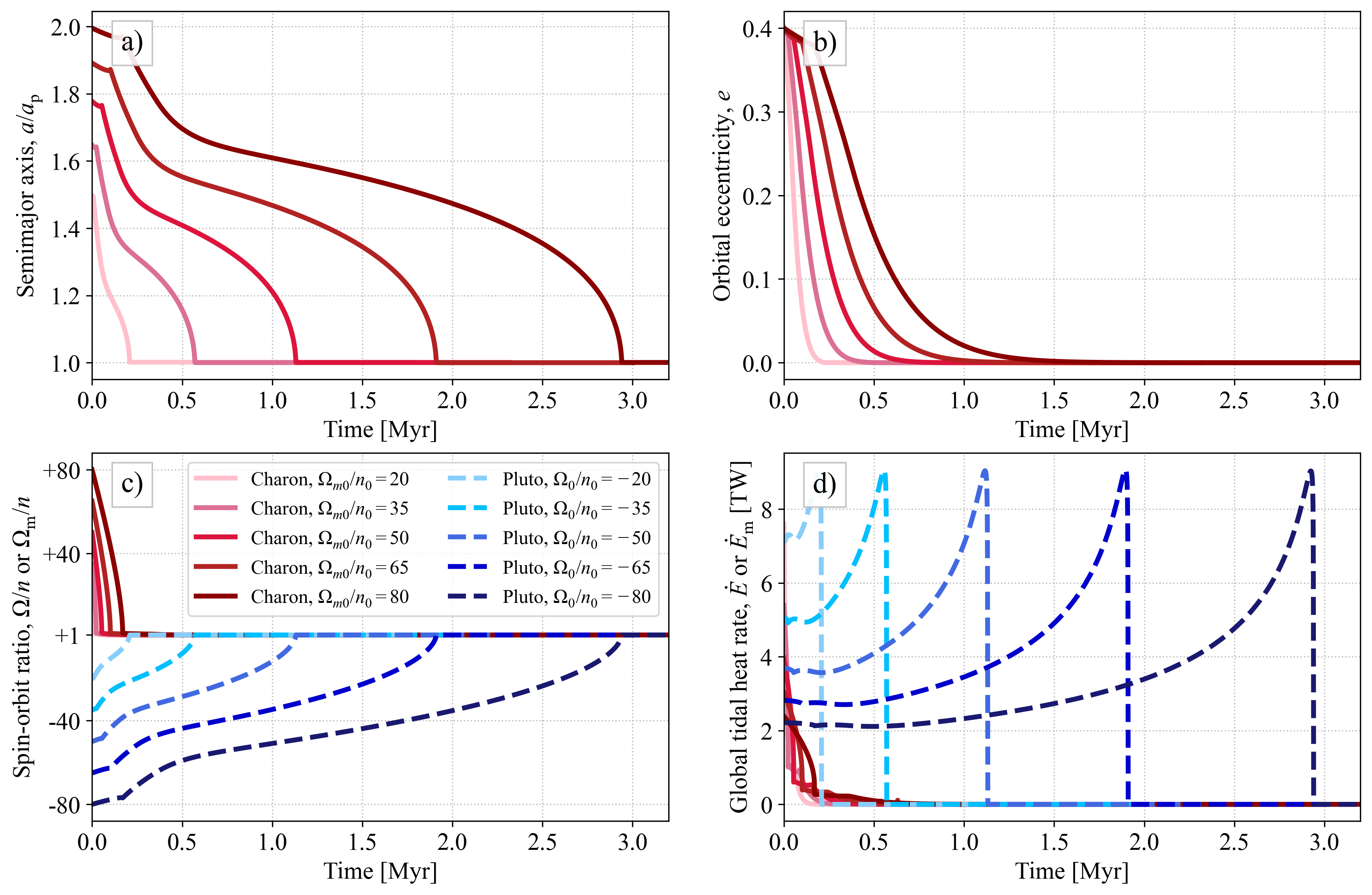}
    \caption{Tidal evolution of the Pluto-Charon system. An initially prograde Pluto and initially retrograde Charon are
considered, with various initial rotation rates and a fixed initial eccentricity $e_0=0.4$. (a) semimajor axis ratio $a/a_{\rm{p}}$ (where $a_{\rm{p}}$ is the present semimajor axis value). (b) eccentricity. (c): spin rates ${\Omega}/n$ (shades of blue) and ${\Omega}_{\rm{m}}/n$ (shades of red). (d) tidally dissipated power, $\dot{E}$ and $\dot{E}_{\rm{m}}$. In accordance with the adopted sign convention, panel (c) shows the spin rate $\Omega$ of an actually prograde Pluto as negative, while the depicted spin $\Omega_{\rm{m}}$ of a retrograde Charon is positive.
    }
    \label{fig:1to1}
\end{figure}

\begin{figure}[h]
  \includegraphics[width=\textwidth]{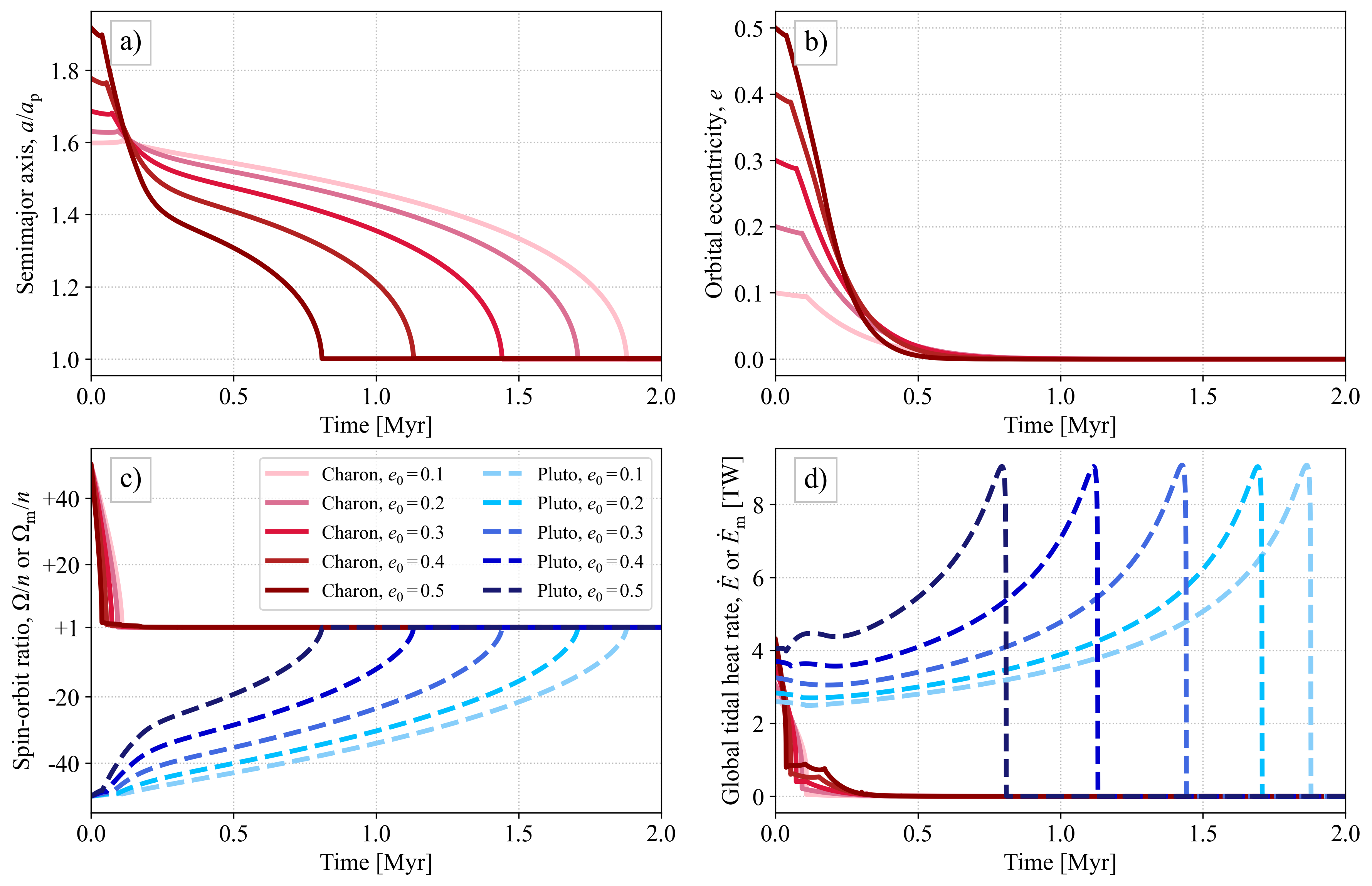}
    \caption{Tidal evolution of the Pluto-Charon system. An initially prograde Pluto and initially retrograde Charon are considered, with various initial eccentricities and fixed initial spin-orbit ratios of $\Omega_0/n_0=-50$ and $\Omega_{\rm{m}0}/n_0=50$. (a) semimajor axis ratio $a/a_{\rm{p}}$ (where $a_{\rm{p}}$ is the present semimajor axis value). (b) eccentricity. (c) spin rates ${\Omega}/n$ (shades of blue) and ${\Omega}_{\rm{m}}/n$ (shades of red). (d) tidally dissipated power $\dot{E}$ and $\dot{E}_{\rm{m}}$.}
    \label{fig:e_variation}
\end{figure}

\begin{figure}[h]
  \includegraphics[width=\textwidth]{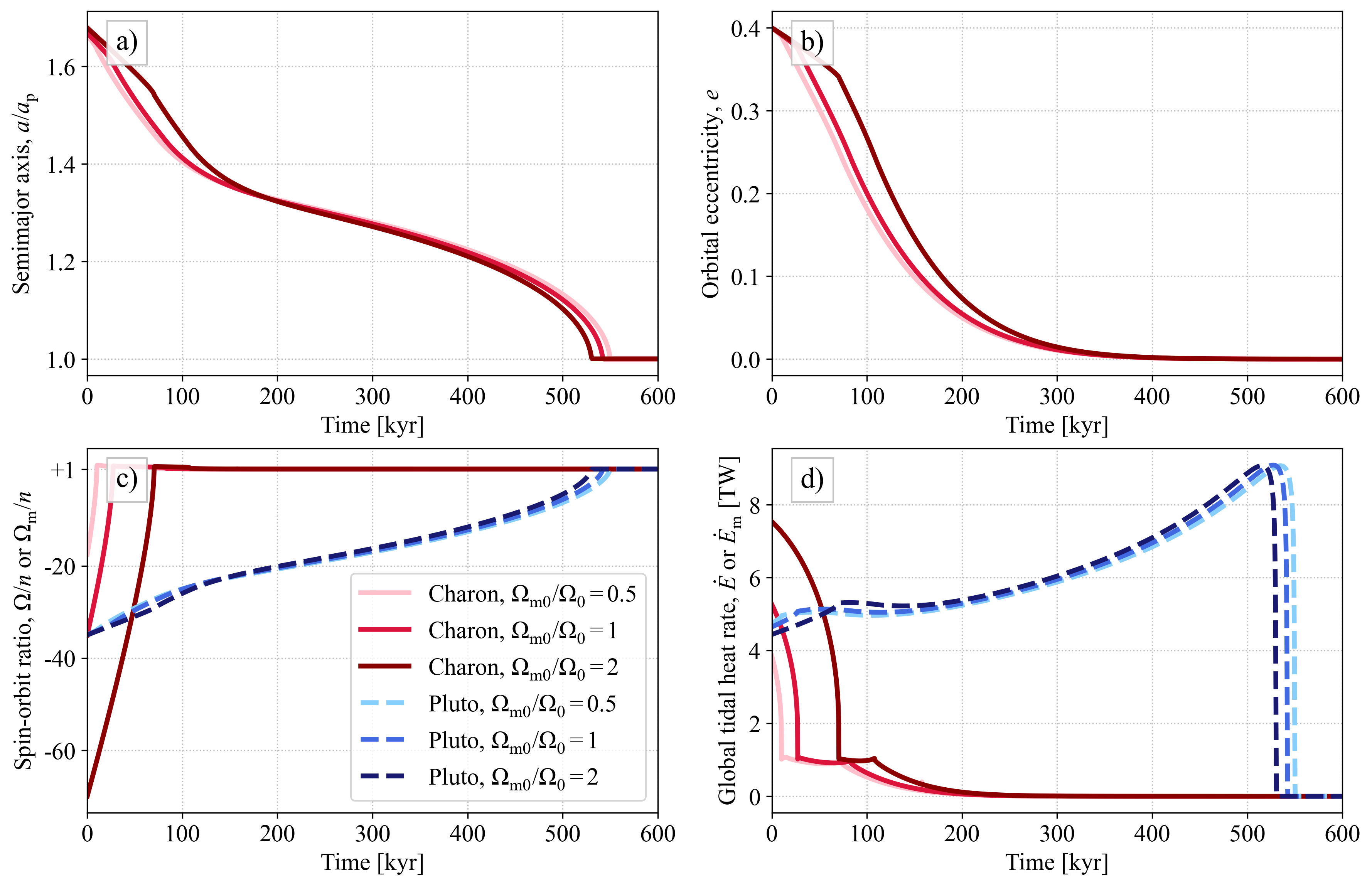}
    \caption{Tidal evolution of the Pluto-Charon system. Initially prograde Pluto and Charon with various initial ratios of rotation rates $\Omega_{\rm{m}0}/\Omega_0$ and the initial eccentricity $e_0=0.4$. (a) semimajor axis ratio $a/a_{\rm{p}}$ (where $a_{\rm{p}}$ is the present semimajor axis value). (b) eccentricity $e$. (c) spin rates ${\Omega}/n$ (shades of blue) and ${\Omega}_{\rm{m}}/n$ (shades of red). (d) tidally dissipated power, $\dot{E}$ and $\dot{E}_{\rm{m}}$.\\
    }
    \label{fig:1to1_negative}
\end{figure}

\subsection{Effect of initial spin rate}

Exploiting the mathematically convenient definition in Section \ref{dynamics} that the orbital motion of Charon is always positive (and prograde within the model), Figure \ref{fig:1to1} illustrates the evolution of an initially prograde Pluto and an initially retrograde Charon for various initial rotation rates and a fixed initial eccentricity of $e_0 = 0.4\s$. \textcolor{black}{This initial configuration corresponds to Case 1 from Table \ref{tab:initial_parameters}, which we shall refer to as our Reference case.} The spin-orbital evolution proceeds relatively quickly and concludes with a doubly synchronised state and a circular orbit within $\unit[250]{kyr}$ for the slowest initial rotation ($14$-hour period for each of the bodies, or $|\Omega_0/n_0|=20$) and within \unit[3]{Myr} for the highest initial rotation rate ($5.4$-hour period, or $|\Omega_0/n_0|=80$).

The evolution begins with a rapid despinning of Charon, completed within $\unit[100-500]{kyr}$. During this initial phase, the semimajor axis $a$ and the eccentricity $e$ experience relatively slow decay because part of the angular momentum originally stored in Charon's rotation is transferred to the orbital motion. The continuous inflow of angular momentum to the orbit compensates for its transfer from the orbit to the rotational motion of Pluto, slowing down Charon's orbital descent. At the same time, Pluto's spin rate slowly `increases' from negative values toward zero.
The despinning of Charon temporarily slows down once Charon's rotation rate approaches spin-orbit resonances 2:1 (i.e., $\Omega_{\rm{m}}/n=2$) and 3:2 (i.e., $\Omega_{\rm{m}}/n=1.5$). However, due to the low ice shell viscosity considered here ($\eta_{\rm{ice}}=\unit[10^{14}]{Pa\;s}\s$; Table \ref{tab:interior_props}) and due to our neglect of the permanent triaxiality, Charon does not get locked into any of these resonances and ultimately despins to the synchronous state ($\Omega_{\rm{m}}/n=1$).\,\footnote{~The reason why a low viscosity is working to prevent a rotator from staying in a higher spin state is explained in detail in \citet[Appendix F]{Mars}. Entrapment into such a state is still possible, probabilistically, for a rotator with a noticeable triaxiality \citep{Makarov2015}~---~which is not the case of Pluto or Charon
 \citep{Nimmo,Nimmo2021}.}

Once the moon attains a synchronous rotation, the orbital eccentricity decay continues at an increased rate, together with the shrinking of the orbit and Pluto's gradual despinning. The eccentricity decreases to zero within $\unit[200]{kyr}$ to $\unit[1.5]{Myr}$ for the individual cases. The rest of the evolution path is marked by the orbital descent of Charon on a circular orbit, and by the despinning, spin reversal, and synchronisation of Pluto's rotation with Charon's mean motion.

Charon is tidally heated only during its relatively short despinning stage (several tens or hundreds of $\unit{kyr}$), and the generated power decreases quickly: it is higher than $\unit[2]{TW}$ only during the few initial $\unit{kyr}$. Pluto, on the other hand, experiences considerable tidal heating during the system's entire spin-orbital evolution. Despite the fast eccentricity decay, the heat generation increases with time as a result of Charon's approach and Pluto's non-synchronous rotation. It ends abruptly upon spin-orbit synchronisation. Due to the dependence of the initial semi-major axis on the initial rotation rate of the partners, the total amount of generated heat is maximal in the case with an initial $5.4$-hour period, the corresponding heating rate exceeding $\unit[2]{TW}$ for $\unit[3]{Myr}$. The peak heating rate in all cases is $\unit[8]{TW}$. If we assume that most of the energy is dissipated in the ice layer \citep[e.g.,][]{henning2014} and if we neglect regional variations in the heating rate, the said value of power will correspond to a maximum specific heating rate of $\unit[2.8\times10^{-9}]{W\; kg^{-1}}$. The specific tidal heating of Charon is of the order of $\unit[10^{-9}-10^{-8}]{W\; kg^{-1}}$ but lasts for a considerably shorter time. \textcolor{black}{Assuming a latent heat of fusion for ice around $\unit[3\times10^{5}]{J\,kg^{-1}}$, and neglecting any heat transport, the predicted maximum tidal heating of Pluto would be able the melt the ice shell only if it operated for at least $\unit[10]{Myr}$, several times longer than is the maximum evolution time scale in Figure \ref{fig:1to1}. We therefore conclude that while the tidal dissipation might have affected the thermal evolution and the rheological properties of the ice layer, and while it might have molten the ice shell locally, it likely was not high enough to cause any large-scale melting of the two bodies \citep[see also][for an example of a tidally-heated moon with a presumably thick frozen hydrosphere]{petricca2025titan}.}

\subsection{Effect of initial eccentricity}

Figure \ref{fig:e_variation} depicts an initially prograde Pluto and an initially retrograde Charon with various initial eccentricities and with initial spin-orbit ratios of $\Omega_0/n_0=-50$ and $\Omega_{\rm{m}0}/n_0=50$. The overall character of the spin-orbital evolution is identical to the Reference case. Changes in the initial orbital eccentricity lead to variations in the time scale of Pluto's despinning and Charon's orbital descent, while keeping the time scale of the eccentricity decay and the moon's despinning almost unaffected.

In particular, increasing the initial eccentricity from $e_0=0.1$ to $e_0=0.5$ decreases the time needed for attaining the present-day configuration approximately by a factor of two. This also influences the period of time for which tidal dissipation represents an important heat source. We note that the total tidally generated heat is defined by the semimajor axis decrease during the evolution and by the slowing down of the rotation. The orbital eccentricity determines what  this semimajor axis decrease will be, and modulates the output (changes the heat rate) as a function of time: at the highest $e_0$, the heating is initially high ($\dot{E}>\unit[4]{TW}$) but short-lived; at the lowest $e_0$, it is initially lower ($\dot{E}\sim\unit[2.5]{TW}$) but acts for twice as much time.

The initial eccentricity also changes the character of Charon's rapid despinning, with higher $e_0$ enabling a weak ``capture" (slowing-down) close to the spin-orbit resonance as high as 5:2 (followed by 2:1, 3:2, and 1:1);  and with lower $e_0$ leading to a direct evolution from the initial rotation to the 1:1 synchronous state.

\subsection{Effect of initial spin-rate ratio}

Finally, in Figure \ref{fig:1to1_negative}, we observe the evolution of Pluto and Charon rotating initially in the same direction, with Charon still being on a retrograde orbit. For the sake of comparison, we prescribe different initial ratios of the two bodies' spin rates $\Omega_{\rm{m}0}/\Omega_0 \in \{0.5, 1, 2\}$. The initial orbital eccentricity is again $e_0=0.4$ and the initial spin-orbit ratio of Pluto is $\Omega_0/n_0=-35$. Due to its relatively small mass, the variations of Charon's initial spin rate affect the overall orbital evolution only negligibly, and the circularisation of the mutual orbit proceeds within $\unit[550]{kyr}$ in all three cases. The only effect that exhibits dependence on Charon's initial spin rate is then, as can be expected, the time scale of the moon's despinning and the magnitude and time scale of its tidal heating.

At the beginning of the spin-orbital evolution, on the scale of tens of $\unit{kyr}$, Charon experienced rapid despinning and reversal of the spin direction. Since the eccentricity decays at a lower rate, Charon's orbit is, at this point, still highly eccentric ($e>0.3$), and the moon cannot be locked into synchronous rotation (1:1 spin-orbit resonance). Therefore, upon reversal, Charon gets temporarily weakly captured close to a 3:2 resonance, from which state it gradually transitions into the 1:1 resonance around $\unit[70-110]{kyr}$.

\subsection{Effect of interior structure and comparison with the case of a tidally receding Charon}

In addition to the three cases described in Table \ref{tab:initial_parameters} and depicted in Figures \ref{fig:1to1} to \ref{fig:1to1_negative}, we tested the effect of ice shell viscosity, the absence of a liquid ocean under Charon's ice shell, and the synchronisation by tidal recession (Scenario~1). The results of these supplementary evolution runs are depicted in  Figures \ref{fig:supp1} - \ref{fig:supp3} in Appendix \ref{A2}.\\

In the first of the additional tests, we reran the Reference case with $\eta_{\rm{ice}}=\unit[10^{16}]{Pa\;s}$ instead of $\unit[10^{14}]{Pa\;s}$. As can be seen in Figure \ref{fig:supp1}, the higher viscosity increases the evolution time scales approximately by a factor of three. Additionally, it influences the evolution trajectory by stabilising Charon's higher spin-orbit resonances. Upon despinning from the initial fast rotation, the moon gets locked in the 7:2 resonance, where it remains for several $\unit{kyr}$ to several tens of $\unit{kyr}$. At lower orbital eccentricities, high spin-orbit resonances disappear; the gradual eccentricity decay thus results in a cascade of resonances that can be maintained over different periods of time, depending on the initial spin rate. The spin-orbit resonances encountered are 3:1, 5:2, 2:1, 3:2, and finally 1:1, which represents the equality of Charon's spin rate and mean motion. In the case with the slowest initial spin rate, $|P_0|=|P_{\rm{m}0}|=\unit[14]{h}$, each of the resonances between 1:1 and 7:2 can be maintained for about $\unit[100]{kyr}$. In the cases with $|P_0|=|P_{\rm{m}0}|<\unit[9]{h}$, Charon may get locked into each of these higher resonances for $\sim\unit[0.5]{Myr}$, and its journey towards 1:1 synchronisation lasts for $\unit[2-3]{Myr}$. Due to the increased ice shell viscosity, the tidal heat generation in either of the bodies is lower than in the Reference case (Figure \ref{fig:1to1}) and does not exceed $\unit[2.3]{TW}$.

If we assume that the viscosity of ice is, again, $\unit[10^{14}]{Pa\;s}$ but the ocean in Charon is fully frozen, we observe a considerable decrease in the moon's despinning rate and tidally generated power (Figure \ref{fig:supp2}). Moreover, if Charon's initial spin rate was relatively slow ($>\unit[8]{h}$), it can get temporarily locked into higher spin-orbit resonances (between 3:2 and 5:2) that can be maintained for several tens of $\unit{kyr}$. The lower efficiency of angular momentum transfer between the companions also results in a slightly shorter timescale for Pluto's despinning, which similarly shrinks the duration of Charon's orbital descent (the orbital evolution is, in this case, predominantly governed by Pluto). \textcolor{black}{As illustrated in Figure \ref{fig:supp2}}, the longest evolution time scale changes from about $\unit[3]{Gyr}$ to $\unit[2.7]{Gyr}$ \textcolor{black}{with respect to the Reference case}. On the other hand, the time scales of Charon's despinning and of the eccentricity decay to $0$ increase and their duration \textcolor{black}{is} the same as for Pluto's despinning and the semimajor axis evolution. The peak tidal heat rate in Pluto then reaches $\unit[10]{TW}$ while the peak tidal heat rate in Charon is always lower than $\unit[0.6]{TW}$, corresponding to a globally averaged heat generation of $\unit[4\times10^{-10}]{W\;kg^{-1}}$ at maximum (for the highest initial spin rate). \textcolor{black}{An initially fully frozen Charon within this scenario is, therefore, unable to attain a subsurface ocean by the sole action of tidal heating.}

For the sake of completeness, we also applied the tidal model to the case of tidal recession (Scenario 1), where the two partners started at the semimajor axis of $a_0 = 0.78\; a_{\rm{p}}\s$, with the initial spin-orbit ratios $\Omega_0/n_0=\Omega_{\rm{m}0}/n_0=13$ and an initial orbital eccentricity $e_0=0.4$. \textcolor{black}{This case is depicted in Figure \ref{fig:supp3}.} The evolution is qualitatively similar to Figure 2 of \citet{Amirs}, with the exception of an initial capture of Charon into a 2:1 resonance, in which it can reside for several hundreds of years. In comparison to the tidal descent (Scenario 2, Figures \ref{fig:1to1} to \ref{fig:supp2}), the time scale of full synchronisation is two orders of magnitude shorter and the tidal heat generation is two orders of magnitude higher, i.e., $\dot{E}\sim\unit[100]{TW}$.

 \section{Conclusions and Outlook}
 \label{summary}

\subsection{Main results}

We opened our exposition with a description of the difficulties faced by giant impact models, and explained how these difficulties may be resolved within the capture scenario, a possibility arising from the study of the dissociation of a double asteroid encountered by Pluto \citep{Agnor, Williams_and_Zugger} and from investigation of chaos-assisted capture \citep{Venus}. We then analysed the orbital evolution of the Pluto–Charon system after Charon’s capture into a Pluto-bound orbit. Assigning to Pluto and Charon initial rotation rates typical of asteroids of comparable size, we deduced that Charon was most probably captured in a highly inclined and retrograde orbit with respect to the Solar-system angular momentum. Charon, while descending, as a result of the action of the tidal torque produced by the tides on Pluto, reversed the rotation of Pluto within a few Myr. \textcolor{black}{With respect to the Solar system angular momentum, Pluto’s initial spin was prograde, while Charon’s initial mean motion was
retrograde. \textcolor{black}{In our notation, this condition (pertaining to the tidal-descent scenario) is indicated by a sign of Pluto's initial rotation rate $\Omega$ that is opposite to Charon's positive definite mean motion $n$ (Figure  \ref{Figure_1}).} Pluto's initial prograde spin is strongly favoured within the pebble accretion scenario of its origin.\,\footnote{~\textcolor{black}{The probability distribution of Pluto's initial rotation directions depends on its formation history. If Pluto predominantly grew by pebble accretion \citep[e.g.][]{Ormel}, its spin would be prograde with high probability \citep[e.g.][]{Takaoka}. If however it
grew through collisions and mergers between dwarf planets, the spin direction could
be distributed almost randomly \citep[e.g.][]{Kokubo_2010}.
In application to Pluto, we would favour pebbles accretion over collisions, because Pluto’s near-perfectly spherical
shape is not immediately reconcilable with major impacts
\citep{plutocharon}.
}
}
}
The initial prograde orbit of Charon is, \textcolor{black}{within the tidal-descent scenario,} less likely, given the narrowness of the parameter space between the loci of stable synchronous states and zero rotation (Figure \ref{Figure_1}). These initial configurations ensured Charon's tidal orbital inward migration, contrary to the commonly assumed concept of outward migration. We studied, both analytically and numerically, the orbital and rotational evolution of Pluto and Charon in the proposed inward-migration regime. Due to the relatively large Charon-to-Pluto mass ratio, this evolution drives the system toward the current tidal synchronisation state across a wide range of parameters.\\

In comparison with \textcolor{black}{the more commonly considered case of tidal recession, the tidally-approaching Pluto-Charon binary experiences substantially weaker tidal heating}. As can be seen in Figure \ref{fig:1to1}d, for example, the values of dissipated power are ranging in the interval $\unit[2-9]{TW}$ for Pluto and up to $\unit[8]{TW}$ for Charon, corresponding to a specific heating rate of the ice shells on the order of $\unit[10^{-9}-10^{-8}]{W\;kg^{-1}}$. This is \textcolor{black}{at least an order of magnitude less than would be required for melting the ice and} roughly $100$ times less than would be generated in the case of outward migration with the same internal structure. Along with lower tidal dissipation, the inward-migration scenario also predicts lower tidal stresses, offering an explanation for the present-day absence of tidally generated fractures \citep{Rhoden}.

Our scenario explains Pluto’s present retrograde rotation in the Solar system, as evidenced by Figures \ref{fig:1to1}--\ref{fig:1to1_negative} (panels (c)), which show various initial spin configurations of Pluto and Charon. In Figures \ref{fig:1to1} and \ref{fig:e_variation}, Charon's initial spin is set retrograde, while in Figure \ref{fig:1to1_negative} it is set prograde, i.e., concordant with that of Pluto.
We found that the initial spin of Charon has little bearing on the character and time scales of the orbital evolution and tidal heating because Charon's contribution to the total angular momentum is relatively small.
In an additional set of cases \textcolor{black}{(Figures \ref{fig:supp1} and \ref{fig:supp2}, Appendix \ref{A2})}, where we increased the ice shell viscosity ($\unit[10^{16}]{Pa\;s}$ instead of $\unit[10^{14}]{Pa\;s}$) and solidified Charon's hydrosphere, we observed that Charon can get temporarily locked into higher spin-orbit resonances (between 3:2 and 7:2), which can be maintained for up to $\unit[0.5]{Myr}$.\\

The existing images of canyons on Charon indicate that it once had an internal ocean \citep{Rhoden2023}. At the same time, simulations of the simultaneous thermal and orbital evolution show that Charon's present-day interior is most likely fully solidified \citep{Amirs}. As proposed by \citet{Rhoden}, the absence of tidally-oriented fractures implies two possibilities for the ocean's fate, which are also applicable in case of Charon's capture. One option is that the ocean, resulting from accretion and radiogenic heating, froze early on and stayed frozen throughout capture and the subsequent circularisation. Another option is that the ocean was sustained even after capture~---~and stayed molten until the end of circularisation. If Charon were fully frozen at the moment of capture, its subsequent re-melting by tides would be highly unlikely. The peak tidal heat rate indicated by our simulations with an oceanless moon is of order  $\unit[10^{-10}]{W\;kg^{-1}}$. Even without the consideration of any heat losses to space, this peak heat rate would need to operate for tens of $\unit{Myr}$ to accumulate an equivalent of the latent heat of melting and to create a liquid ocean. Yet, according to our models, the evolution was concluded within $\unit[3]{Myr}$. On the other hand, if Charon had entered the tidal circularisation stage as partially molten, the increased tidal heating might have helped it to sustain its subsurface ocean until the end of tidal evolution. A clearer picture of the ocean's lifetime and later disappearance would be provided by coupling our spin-orbital model with a thermal evolution model, as considered elsewhere e.g.,~\citep{Robuchon,Amirs}.\\

In conclusion, we believe that a synchronisation of Pluto and Charon by inward orbital migration of an initially retrograde moon provides an internally consistent explanation for the present-day retrograde rotation of Pluto as well as for the absence of tidally-generated tectonic patterns on its surface, which is present on tidally receding ocean-bearing icy satellites of Jupiter and Saturn \citep[e.g.,][]{matsuyama2008,kattenhorn2009}. Since the attainment of a retrograde, widely separated, and similarly dense moon is not easily explained by giant-impact models, these results indicate that a capture scenario is indeed viable.

\subsection{Topics for further research}

As we mentioned above, the capture process should either avoid remelting Charon or circularise its orbit before (re)freezing the interior. Otherwise, the large stresses from ocean freezing would have generated fractures with orientations guided by the diurnal tidal stress, a pattern not seen on Charon's surface \citep{Rhoden}.
While within our scenario the rate of tidal dissipation is two orders of magnitude lower than in the outward-migration case, it is still intense. On the other hand, it lasts for a period shorter than \unit[10]{Myr}. The question of Pluto's and Charon's thermal evolution depends on a variety of factors, like the thermophysical properties of the ice or the distribution of tidal heating and possible partial melting, and needs to be addressed separately.

We also note that in this work we only study the tidal evolution of the Pluto-Charon binary in a coplanar configuration, without assuming the evolution of Charon's obliquity and inclination with respect to a reference plane. This is certainly a simplification and the full evolution might reveal additional dynamical effects. Likewise, we start the numerical calculations in an initial state where Charon has already attained a Pluto-bound orbit. It remains to be explored under what conditions the hypothetical dissociation of a binary asteroid \citep{Agnor, Williams_and_Zugger} produces a Charon-like moon orbiting Pluto and whether this process alone may lead to any considerable dissipation that would affect the interior structures of the bodies.

 One more topic for study could be the fracturing of Charon by tidally driven stresses whose magnitude is maximal at large eccentricities. As was mentioned in Section \ref{Introduction}, with reference to \citet{Rhoden}, one might expect to observe on an outward-migrated Charon a picture of oriented fractures similar to the pattern existing on Europa, Enceladus and other large icy moons. No such picture, however, was found on Charon in {\it Ibid.} It therefore may be worth extending these authors' research to a Charon approaching the present orbit from outside. In this case, the eccentricity was assuming  higher values during the period of largest, not closest separation between the partners, the resulting tidal stresses therefore being weaker by an order or two of magnitude because the perturbing tidal potential scales as $a^{-3}$. We therefore would expect no appreciable tidal fracturing of a captured Charon. Still, a dedicated study in the spirit of {\it Ibid.} is required to establish a final verdict on this.

Like Charon, Pluto's small moons
 ~---~Styx, Nix, Kerberos, and Hydra~---~
are in near-equatorial low-eccentricity orbits. These orbits are close to integer mean-motion resonances with Charon's orbit, none of those resonances being exact. Assuming that Charon was born through a giant impact,  \citet{Ward_and_Canup_2006} proposed that the orbits of Nix and Hydra could have been driven outward by resonant interactions with Charon during its tidal orbital expansion. Naturally, the question now emerges as to how tidal contraction of Charon's orbit could have influenced the small satellites, assuming that their emergence predates that of Charon. Additionally, the tidal capture scenario reopens the question of the moons' origin and orbital evolution that lead to their present-day coplanar configuration.

Finally, while it is possible that the inward-migration of a retrograde moon reversed the rotation of Venus \citep{Venus}, it remains to be explored if this mechanism could have been instrumental also in transforming Uranus' obliquity on the ecliptic through the agency of a moon that had been captured in a near-polar orbit. Uranus' rings and  their shepherds may contain that moon's remnants.

\begin{acknowledgments}
 The authors would like to thank the Editor and the two anonymous Reviewers for their valuable comments and suggestions that helped to improve the manuscript.
 One of the authors (ME) is indebted to Sota Arakawa, Alyssa Rhoden and Francis Nimmo for helpful consultations.
 The authors are also grateful to Darin Ragozzine for the useful comments and references that he kindly provided.
 MW acknowledges the support from Czech Science Foundation grant nr. 23-06513I.
\end{acknowledgments}

\begin{contribution}
 VM and ME came up with the initial research concept.
 ME was responsible for writing most of Sections \ref{Introduction}, \ref{histories} and \ref{summary}, as well as for writing Section \ref{dynamics} and Appendix \ref{A1}.
 VM provided validation of the analytical theory in Sections \ref{dynamics} and \ref{histories}, and participated in editing the manuscript.
 MW developed the software, ran simulations, plotted the results, wrote most of Sections \ref{numerics} and \ref{runs}, contributed to Sections \ref{histories}  and \ref{summary}, and participated in editing the manuscript.
 YG ran  preliminary qualitative simulations, participated in analysing the errors and interpreting the plots. She also contributed to Sections \ref{histories} and \ref{numerics},  participated in editing the manuscript, and submitted it to the journal.
 AB contributed to interpreting the results, and to writing Sections \ref{numerics} and \ref{runs}, and participated in editing the manuscript.
 AK contributed to interpreting the results and to writing Sections \ref{Introduction}, \ref{numerics} and \ref{runs}, and participated in editing the manuscript.

\end{contribution}

\appendix

\section{Derivation of equation (1)}
\label{A1}

 With $\Omega$ and $\Omega_{\rm{m}}$ being the rotation rates of the partners, $M$ and $M_{\rm{m}}$ their masses, $R$ and $R_{\rm{m}}$ their radii, $C=\xi M R^2$ and $C_{\rm{m}}=\xi_{\rm{m}} M_{\rm{m}} R_{\rm{m}}^2$ their moments of inertia,
  the polar (perpendicular to the orbit) component of the angular momentum vector, in the centre of mass system, reads \citep[eqn 132]{book}:
 \ba
 H & = & \frac{M\s M_{\rm{m}}}{M +\s M_{\rm{m}}}\,\sqrt{G\s (M +\s M_{\rm{m}} )\,a\,(1\,-\,e^2)\,}
 \nonumber\\
 \label{en}
 \label{route}
 \label{to}\\
 \nonumber
 & + & C \,{ \Omega }\s+\s C_{\rm{m}}\s\Omega_{\rm{m}}\s+\s O(i^2)\s+\s O(i_{\rm{m}}^2)\,\;.
 \ea
Similarly to other standard texts, the cited equation (132) from \citet{book} gives the angular momentum per unit mass,
 $\,\sqrt{G\s (M +\s M_{\rm{m}} )\,a\,(1\,-\,e^2)\,}\,$. To obtain the physical angular momentum, it must be multiplied by the reduced mass $\,{M\s M_{\rm{m}}}/(M +\s M_{\rm{m}})\,$. Hence the above expression (\ref{route}).

 In neglect of the planet's and moon's squared obliquities on orbit, $i^2$ and $i_{\rm{m}}^2$, we write the time derivative of $H$ as
 \ba
 0 &=& \frac{M \,M_{\rm{m}}}{M + M_{\rm{m}}}\;\sqrt{G\,(M + M_{\rm{m}})\,(1-e^2)\,}\;\frac{d}{dt}\s a^{1/2}
 \nonumber\\
 \nonumber\\
   &-& \frac{M \,M_{\rm{m}}}{M + M_{\rm{m}}}\;\sqrt{G\,(M + M_{\rm{m}})\,a\,}\,\frac{e}{\sqrt{1-e^2}}\,\frac{de}{dt}
 \label{L}\\
 \nonumber\\
 \nonumber
   &+& C \,{\stackrel{\bf\centerdot}{\Omega\s}} \,+\,C_{\rm{m}} \,{\stackrel{\bf\centerdot}{\Omega\s}}_m
 \;\;.
 \ea
 Casting $\s n=\sqrt{G(M+M_{\rm{m}})\, a^{-3}\,}\,$ into $\,a^{\s 1/2}=(G\s(M+M_{\rm{m}}))^{1/6}\s n^{-1/3}\s$, we derive the equality
 \ba
 \nonumber
 \frac{d}{dt}\s a^{\s 1/2} &=&-\,\frac{1}{3}\,(G(M+M_{\rm{m}}))^{1/6}\s n^{-4/3}\, {\stackrel{\bf\centerdot}{n\s}}\\
 \label{}\\
 &=&-\,\frac{1}{3}\,(G(M+M_{\rm{m}}))^{-1/2} \s a^{2} \,{\stackrel{\bf\centerdot}{n\s}}\;\;,
 \nonumber
 \ea
 insertion whereof into equation (\ref{L}) entails
 \ba
 \nonumber
 0\;= &-&\frac{1}{3}\,\frac{M \,M_{\rm{m}}}{M + M_{\rm{m}}}\;a^2\,{\stackrel{\bf\centerdot}{n\s}}\,\sqrt{1-e^2}\\
 \nonumber\\
 &-& \frac{M \,M_{\rm{m}}}{M + M_{\rm{m}}}\;n\;a^2\;\frac{e\;\dot{e}}{\sqrt{1-e^2}}
 \label{put}\\
 &+&
 \s \xi\s M\s R^{\s 2} \,{\stackrel{\bf\centerdot}{\Omega\s}}
 \s+\s\xi_m\s M_{\rm{m}}\s R_{\rm{m}}^{\s 2} \,{\stackrel{\bf\centerdot}{\Omega\s}}_{\rm{m}}
 \;\;.
 \nonumber
 \ea

 This can also be written as
 \ba
 {\stackrel{\bf\centerdot}{\Omega}} = \frac{1}{\xi}\;\frac{M_{\rm{m}}}{M + M_{\rm{m}}}\;n\;\left(\frac{a}{R}\right)^2\;\frac{e\;\dot{e}}{\sqrt{1-e^2}}\;+ \qquad\qquad\qquad\qquad\qquad\qquad\qquad\qquad\qquad\qquad
 \nonumber\\
 \label{relation}\\
 \nonumber
 \frac{  {\stackrel{\bf\centerdot}{n\s}}  }{3\s\xi}\,\frac{M_{\rm{m}}}{M + M_{\rm{m}}}\left(\frac{a}{R}\right)^2
 \sqrt{1-e^2}
 \; \left[1
 -\,3\s\xi_m\s \frac{M+M_{\rm{m}}}{M}\s \left(\!\frac{R_{\rm{m}}}{a}\!\right)^{\s 2}\frac{1}{\sqrt{1-e^2}}\,\frac{\,  {\stackrel{\bf\centerdot}{\Omega}}_{\rm{m}} \, }{  {\stackrel{\bf\centerdot}{n\s}}  }
 \right]\;.\;\;\;
 \ea
Proportional to $\xi_m
\s$, the second term in square brackets emerges due to the spin angular momentum of the moon. To drop this term is the same as to neglect the input from the spin of the moon into the angular-momentum balance.
 Assuming that $3\s\xi_m\simeq 1$, and employing the values $M = 1.302 \times 10^{\s 22}~\mbox{kg}\s$, $\,M_{\rm{m}} = 1.586 \times 10^{\s 21}~\mbox{kg}\s$, $R_{\rm{m}} = 6.060\times 10^{5}$~m, and $\,a=1.960\times 10^{7}$ m, we observe that for Pluto and Charon that term is
 \ba
 \frac{\textstyle 1.07\times 10^{-3}}{\textstyle\sqrt{1-e^2}}\,\frac{\,  {\stackrel{\bf\centerdot}{\Omega}}_{\rm{m}} \, }{  {\stackrel{\bf\centerdot}{n\s}}  }\,\;,
 \label{neglected}
 \ea
 where the coefficient $\,1.07\times 10^{-3}/\sqrt{1-e^2}\,$
  is below $0.13\%$ for $e<0.4$ and barely exceeds $0.2\%$ for $e=0.85$.

 In the limit of the moon's spin getting synchronous with the orbit,  we set $\s{\Omega\s}_{\rm{m}} = {n}\s$ and ${\,  {\stackrel{\bf\centerdot}{\Omega}}_{\rm{m}} \, }=\s{  {\stackrel{\bf\centerdot}{n\s}}  }$. In this situation, the second term in square brackets may be omitted, and formula (\ref{relation}) becomes, to a percent precision level,~simply
 \begin{subequations}
 \ba
 \frac{{\stackrel{\bf\centerdot}{\Omega\s}}}{{\stackrel{\bf\centerdot}{n\s}}}
 &=&
 \frac{1}{3\s\xi}\,\frac{M_{\rm{m}}}{M + M_{\rm{m}}}\left(\frac{a}{R}\right)^2 \sqrt{1-e^2}\left[1\,+\,3\; \frac{n\;e}{1-e^2}\; \frac{\,{\stackrel{\bf\centerdot}{e\s}}\,}{{\stackrel{\bf\centerdot}{n\s}}} \right]
 \label{}\\
 \nonumber\\
 &=&
 \frac{1}{3\s\xi}\,\frac{M_{\rm{m}}}{M + M_{\rm{m}}}\left(\frac{a}{R}\right)^2 \sqrt{1-e^2}\left[1\,-\,2\; \frac{a\;e}{1-e^2}\; \frac{\,{\stackrel{\bf\centerdot}{e\s}}\,}{{\stackrel{\bf\centerdot}{a\s}}} \right]\;\,,
 \label{}
 \ea
 \label{25}
 \end{subequations}
 where we recalled that $\,3\s{\textstyle n}/{\textstyle{\stackrel{\bf\centerdot}{n\s}}}\s=\s-\s 2\s{\textstyle a  }/{\textstyle  {\stackrel{\bf\centerdot}{a\s}} }\,$.

 As well known (see, e.g. \citeauthor{BoueEfroimsky2019} \citeyear{BoueEfroimsky2019}), the tidal rate ${\s{\stackrel{\bf\centerdot}{e\s}}\s}$ scales as $e^1$. The rate ${\s{\stackrel{\bf\centerdot}{a\s}}}\s$ scales as $e^0$ insofar as the spin of at least one partner remains nonsynchronous~---~which is the setting in which equation (\ref{25}) is used and in which we therefore are interested. For this setting, we thus finally arrive at
 \ba
 \frac{{\stackrel{\bf\centerdot}{\Omega\s}}}{{\stackrel{\bf\centerdot}{n\s}}}
 \,=\,
 \frac{1}{3\s\xi}\,\frac{M_{\rm{m}}}{M + M_{\rm{m}}}\left(\frac{a}{R}\right)^2 \,+\,O(e^2)\;\,.
 \label{derivation}
 \ea
The partner last to synchronise its rotation is Pluto, because a less massive partner synchronises its rotation first. This can be understood from the comparison of formulae for the partners' despinning timescales, provided in \citet[Appendix D1]{Mars}. Our expression (\ref{derivation}) is thus valid over the final part of evolution, when Charon is synchronised, while Pluto is not yet. In the limit of vanishing eccentricity, the expression becomes exact. It however should be noted that at an earlier stage of evolution, when
Charon's spin is not yet synchronous and the above-neglected term (\ref{neglected}) is not small, this expression is only an approximation, even for small eccentricities.

\section{Supplementary runs}
\label{A2}

\begin{figure}[h]
  \includegraphics[width=\textwidth]{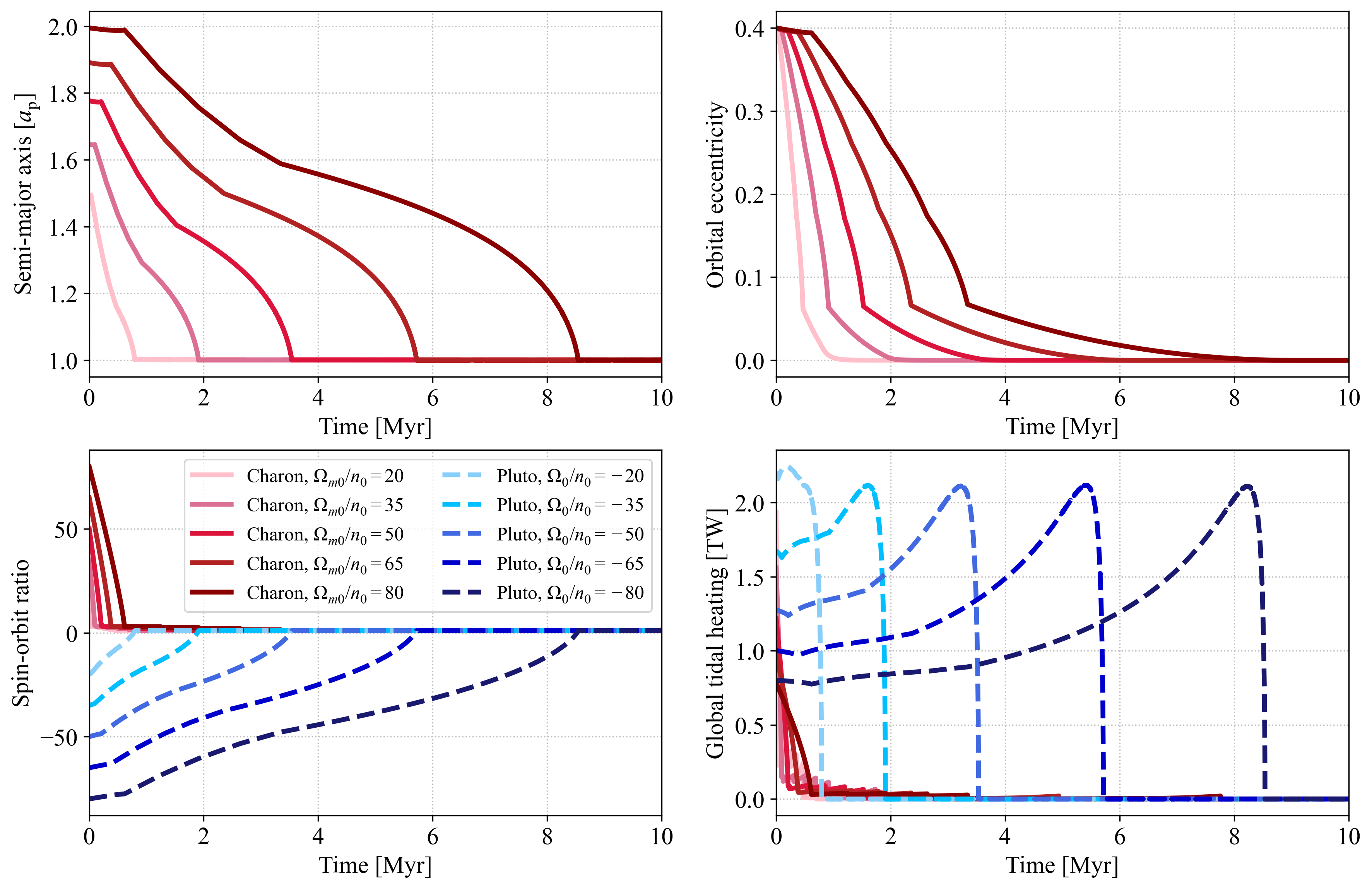}
    \caption{Same as Figure \ref{fig:1to1} but for a higher ice viscosity, $\eta_{\rm{ice}}=\unit[10^{16}]{Pa\; s}$.}
    \label{fig:supp1}
\end{figure}

\begin{figure}[h]
  \includegraphics[width=\textwidth]{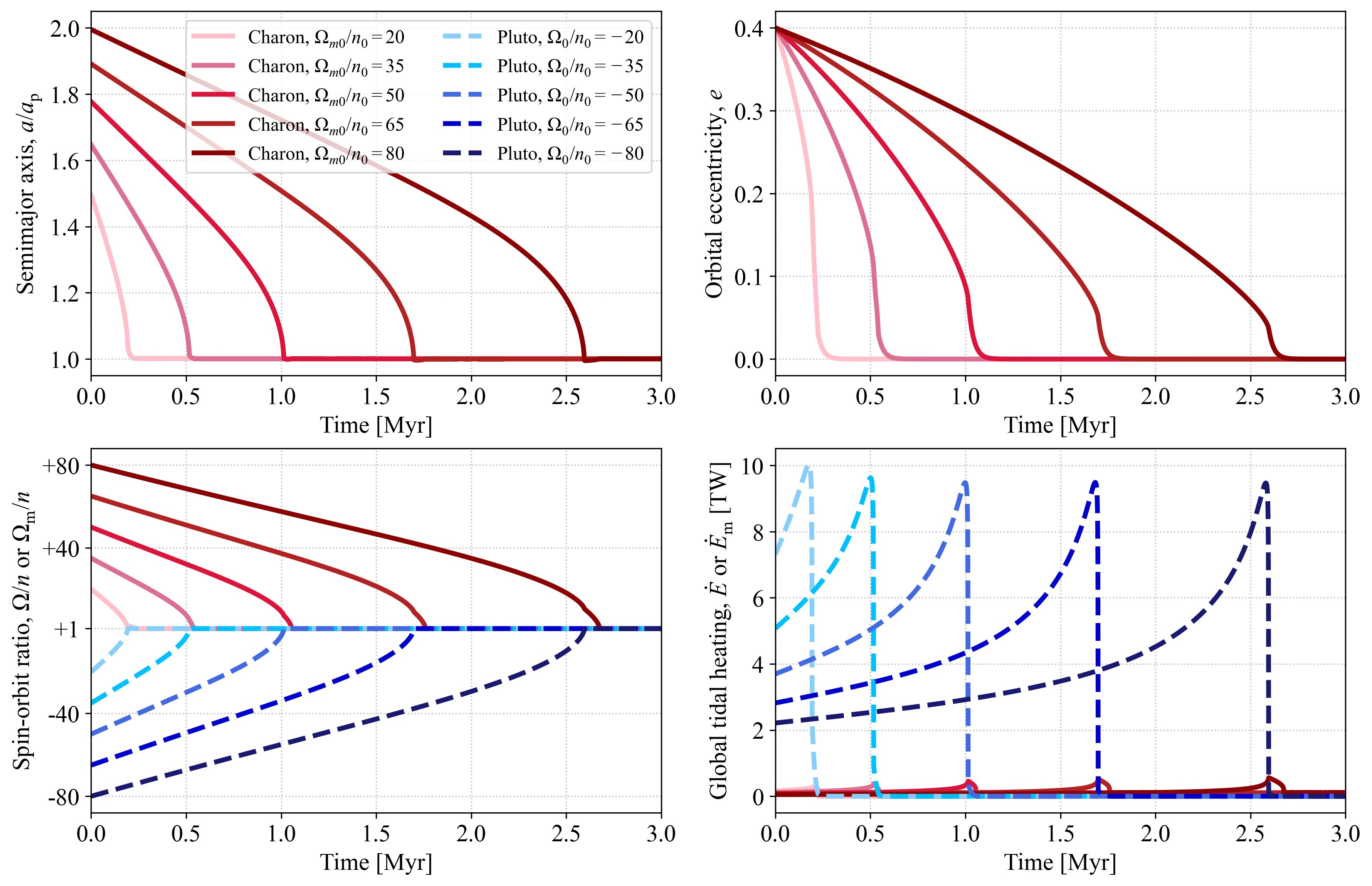}
    \caption{Same as Figure \ref{fig:1to1} but for a fully frozen Charon (the entire hydrosphere consists of ice with viscosity $\eta_{\rm{ice}}=\unit[10^{14}]{Pa\; s}$).}
    \label{fig:supp2}
\end{figure}

\begin{figure}[h]
  \includegraphics[width=\textwidth]{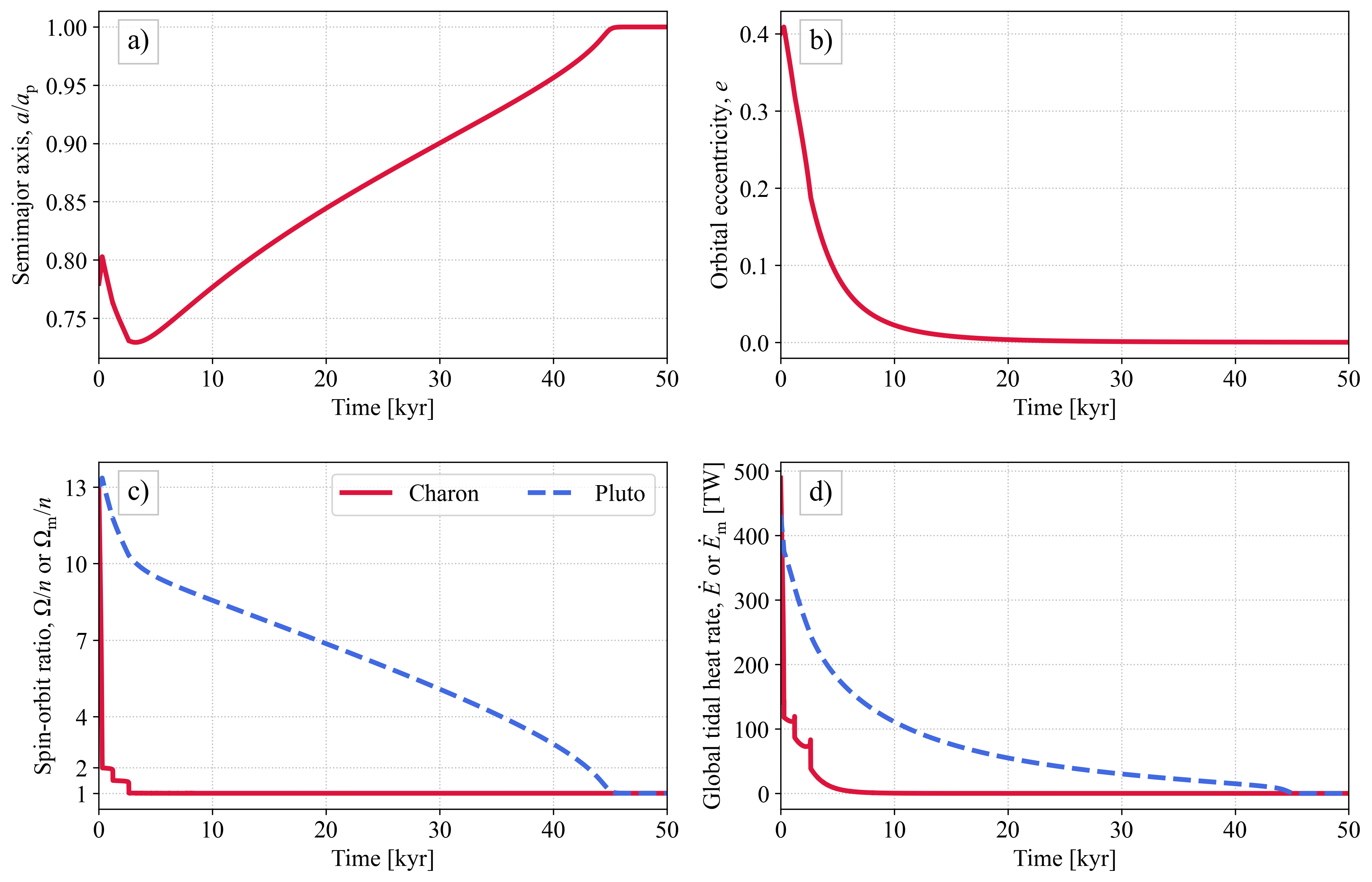}
    \caption{Orbital and rotational evolution and tidal heating of Pluto and Charon experiencing tidal ascent (Scenario 1).}
    \label{fig:supp3}
\end{figure}

Figures \ref{fig:supp1}, \ref{fig:supp2} and \ref{fig:supp3} present  the evolution paths taken by Pluto and Charon in the additional numerical studies, as described in the last four paragraphs of Section \ref{runs}.

 \clearpage
\bibliography{references}{}
\bibliographystyle{aasjournal}

\end{document}